\documentclass[%
twocolumn,
amsmath,amssymb,
aps,
prx,
]{revtex4-1}

\usepackage{graphicx,epsfig,epstopdf}
\usepackage{amsmath}
\usepackage{color}
\usepackage{caption}
\usepackage{subfigure}
\usepackage{array}
\usepackage{tabularx}
\usepackage{multirow}
\usepackage{gensymb}
\newcolumntype{L}[1]{>{\raggedright\arraybackslash}p{#1}}
\newcolumntype{C}[1]{>{\centering\arraybackslash}p{#1}}
\newcolumntype{M}[1]{>{\centering\arraybackslash}m{#1}}

\def\e{\begin{equation}}
\def\f{\end{equation}}
\def\_#1{{\bf #1}}
\def\.{\cdot}

\def\Re{{\rm Re\mit}}

\begin{document}

\title{Flat Engineered Multi-Channel Reflectors}

\author{
V.~S.~Asadchy$^{1,2}$, A.~D\'{i}az-Rubio$^{1}$, S.~N.~Tcvetkova$^{1}$, D.-H.~Kwon$^{1,3}$, A.~Elsakka$^{1}$, M.~Albooyeh$^{1,4}$,  and   S.~A.~Tretyakov$^{1}$
}
 
\affiliation{$^1$Department of Electronics and Nanoengineering\\ 
Aalto University, P.~O.~Box~15500, FI-00076 Aalto, Finland\\
$^2$Department of General Physics, Francisk Skorina Gomel State University, 246019 Gomel, Belarus\\
{$^3$Department of Electrical and Computer Engineering,
University of Massachusetts Amherst, Amherst, MA 01002, USA}\\
$^4$Department of Electrical Engineering and Computer Science\\
University of California, Irvine, CA 92617, USA}

\begin{abstract}

Recent advances in engineered gradient metasurfaces have enabled unprecedented opportunities for light manipulation using optically thin sheets, such as anomalous refraction, reflection, or focusing of an incident beam. Here we introduce a concept of multi-channel functional metasurfaces, which are able to control incoming and outgoing waves in a number of propagation directions simultaneously. In particular, we reveal a possibility to  engineer multi-channel reflectors. Under the assumption of reciprocity and energy conservation, we find that there exist three basic functionalities of such reflectors: Specular, anomalous, and retro reflections. Multi-channel response of a general flat reflector can be described by a combination of these functionalities. 
To demonstrate the potential of the introduced concept, we design and experimentally test three different multi-channel reflectors: Three- and five-channel retro-reflectors and a three-channel power splitter. Furthermore, by extending the concept to reflectors supporting higher-order Floquet harmonics, we forecast the emergence of other multiple-channel flat devices, such as isolating mirrors, complex splitters, and multi-functional gratings.

\end{abstract}

\maketitle

\section{Introduction}

Recently, it was shown that thin composite layers (called \emph{metasurfaces}) can  operate as effective tools for controlling  and transforming electromagnetic waves, see review papers  
\cite{Gly,Hol_review,Shalaev_review,Capasso_review,my_review,Mohammad324}. A number of fascinating and unique functionalities have been   realized with the use of thin inhomogeneous layers, such as anomalous refraction \cite{capasso}, reflection \cite{sun,bozh1}, focusing \cite{capassonew}, polarization transformation \cite{Niemi}, perfect absorption \cite{absor11}, and more \cite{Gly}.
In particular, significant progress has been achieved in controlling reflections of plane waves using structured surfaces.

Methods for engineering single-channel  reflections (when a single plane wave is reflected into another single plane wave) are known. 
According to the reflection law, light impinging on a flat and smooth (roughness is negligible at the wavelength scale) mirror is reflected into the specular direction.  It is simple to show that if there are induced electric or magnetic surface currents at the reflecting boundary, the conventional reflection law, in general, does not hold when the surface properties (e.g., surface impedance) smoothly vary within the wavelength scale. Proper engineering of the induced  surface current gradients enables reflection of the incident light in a direction different from the specular one. This approach applied to metasurfaces was exploited recently in \cite{capasso} and formulated as the generalized law of reflection. It was shown that the desired current gradient can be engineered using a metasurface realized as arrays of  specifically designed sub-wavelength antennas.  A large variety of metasurface designs for reflection control with different power efficiency levels were reported (e.g., \cite{sun,bozh1,bozh2,mosall2,nanotechnology,prx4,veysi,functional,li}). More recently, it was shown that  parasitic reflections in undesired directions, inevitable in the designs based on the generalized reflection law \cite{capasso}, can be removed \cite{synthesis,last} via engineering spatial dispersion in metasurfaces \cite{ana}.

While the recently developed anomalously reflecting metasurfaces greatly extend the functionalities of  conventional optical components such as blazed gratings \cite{handbook}, all these devices   are designed  to control and manipulate incident  fields of only one specific configuration. For example, metasurface reflectors \cite{sun,bozh1} and blazed gratings \cite{grating1,grating2,retro1} refract/reflect a plane wave incident from a specific direction  into another plane wave propagating in the desired direction. 

Recently, a possibility of multi-functional performance  using several parallel metasurfaces each performing its function at its operational frequency was considered in \cite{multi}. Using a single metasurface, it is in principle possible to satisfy boundary conditions for more than one set of incident/reflected/transmitted waves if one assumes that the surface is characterized by a general bianisotropic set of surface susceptibilities~\cite{karim}. However, mathematical solutions for the required susceptibilities may lead to active, nonreciprocal or physically unrealizable parameter values~\cite{karim}.

In this paper, we propose flat metasurfaces with engineered response to excitations by plane waves coming from several different directions. 
As a particular conceptual example  of realizable flat multi-channel devices, here we study multi-channel reflectors: Lossless reciprocal flat  reflectors capable of  simultaneous reflection control  from and into several directions in space. 
We explore  the possible functionalities of general  surface-modulated flat reflectors. To this end, we  characterize  an arbitrary flat periodically modulated reflector using Floquet harmonics as a multi-channel system and inspect all allowed reflection scenarios under the assumption of lossless and reciprocal response. We analyze the reflector response in the framework of  conventional scattering matrix notations.  We find that in the case of periodic reflectors described by a $3 \times 3$ scattering matrix (three-channel reflectors), there are devices with three basic functionalities: General specular reflectors, anomalous mirrors, and three-channel retro-reflectors. Moreover, response of an arbitrary multi-channel ($N \times N$) reflector can be described by a combination of these basic functionalities. 
 Although this general classification reveals all possible functionalities available with periodic reflectors, it does not provide a recipe for designing a metasurface with specific properties.  To this end,   exploiting the general surface impedance model \cite{synthesis}, we rigorously design a three-channel retro-reflector and experimentally verify its electromagnetic response. 
Furthermore, we synthesize two  other  multi-channel devices: A   beam splitter that is matched to the normally incident wave and a five-channel retro-reflector. 
We expect that the developed theory and realizations of multi-channel flat reflectors  lead to more sophisticated  thin and flat $N$-channel metadevices, such as power dividers, directional couplers, interferometers or multi-channel filters for a broad range of frequencies.


\section{Multi-channel paradigm of flat reflectors}
\subsection{Definitions and notations}
In general, reflection from planar surfaces can be controlled by surface structuring either on the wavelength scale (conventional diffraction gratings) or on the sub-wavelength scale (metasurface-based gratings). Operational principles of metasurface-based gratings differ from those of conventional blazed gratings \cite{comparison}. The former   rely on proper phase gradient formed by sub-wavelength scatterers, while the latter   operate due to constructive interference of the rays reflected/refracted from different grooves. Planar topology of  metasurface gratings is an important advantage in fabrication and in some applications. Here we explore the design flexibility offered by the metasurface paradigm.

\begin{figure}[h]
	\centering
\hspace{-0.2cm}	\subfigure[]{
		\epsfig{file=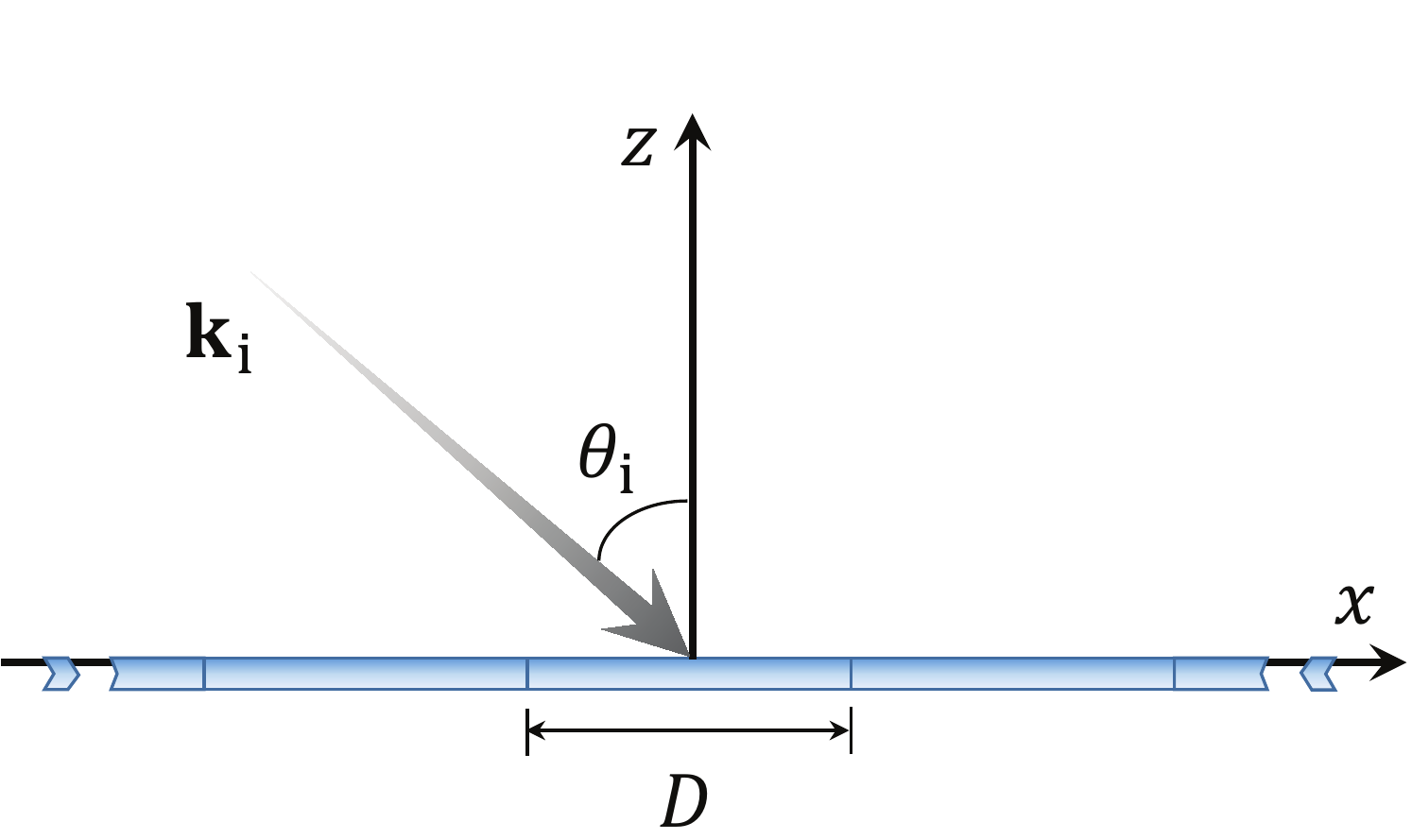, width=0.48\columnwidth}  
		\label{fig1a} }
\hspace{-0.3cm}	\subfigure[]{
		\epsfig{file=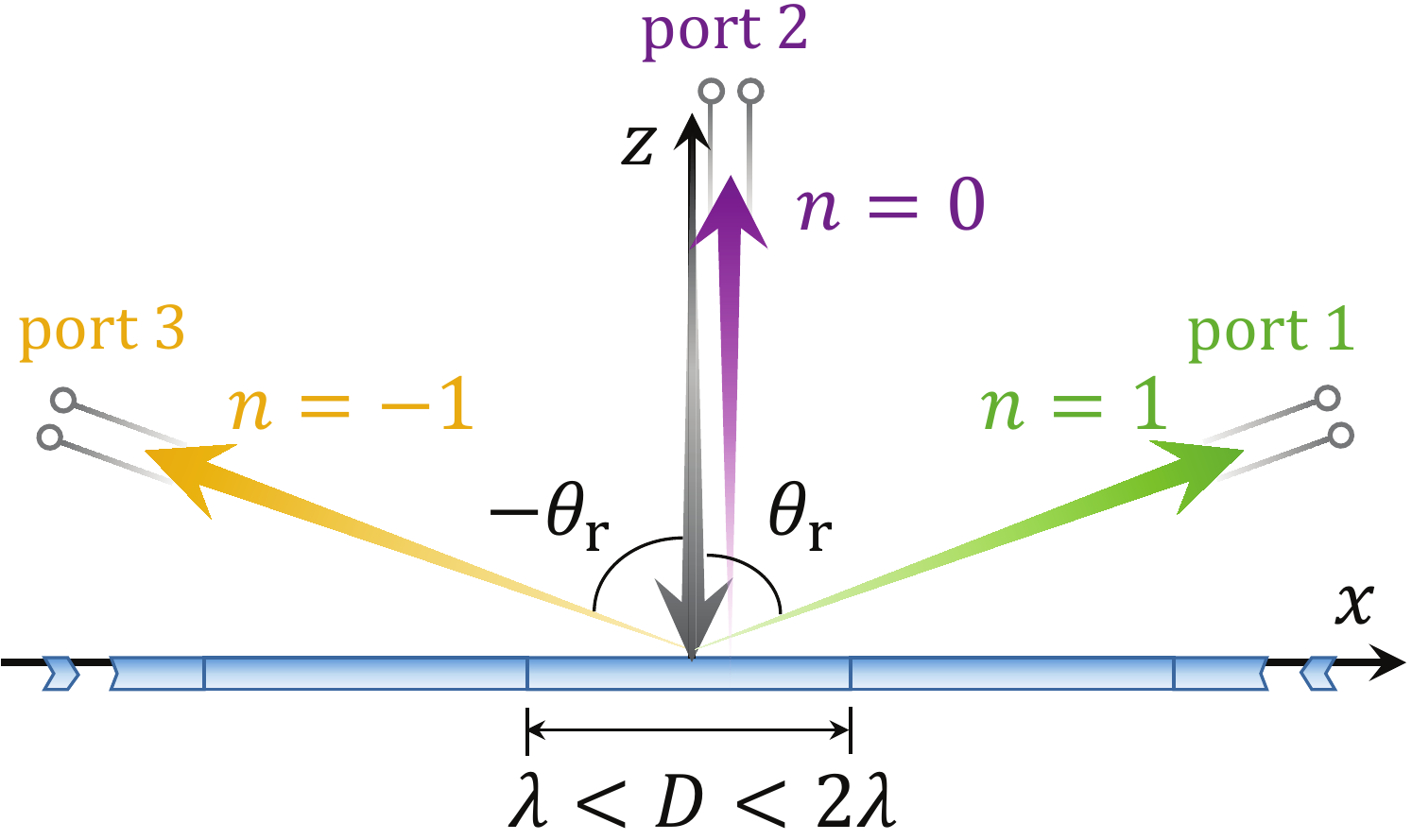, width=0.48\columnwidth} 
		\label{fig1b} }  
         \caption{(a) Illustration of a periodic metasurface illuminated by a plane wave impinging from an angle $\theta_{\rm i}$. (b) A periodic metasurface illuminated at $\theta_{\rm i}=0^\circ$. The three propagating harmonics of the metasurface are analogous to a three-port network. 
	}\label{fig1}
\end{figure} 
Let us consider a periodic metasurface in free space illuminated by a plane wave at an angle $\theta_{\rm i}$ as shown in Fig.~\ref{fig1a}. 
Reflection from a periodic structure, in general, can be represented as interference of the infinite number of propagating and evanescent plane waves (Floquet harmonics). Here,  fully reflecting surfaces (no transmission through the metasurface) are studied. The tangential wavenumber $k_{{\rm r} x}$ of a reflected harmonic of order $n$   is related to the incident wave wavenumber $k_{\rm i}$ and to the period of the structure $D$ as
$k_{{\rm r} x}=k_{{\rm i} }\sin{\theta_{\rm i}}+2\pi n/D$. The corresponding normal wavenumber of the $n$-th harmonic $k_{{\rm r} z}=\sqrt{k_{{\rm i}}^2-k_{{\rm r} x}^2}$ indicates whether it is a propagating or evanescent wave. Figure~{\ref{fig2a} shows the normal component of the reflected wave vector of all possible propagating modes from a flat reflector (for simplicity the incident angle $\theta_{\rm i}=0^\circ$ is assumed) with respect to the period $D$. The reflection angle of the corresponding harmonics $\theta_{\rm r}=\arcsin(k_{{\rm r} x}/k_{{\rm i}})$ is shown in Fig.~\ref{fig2b}. 
\begin{figure}[h]
	\centering
	\subfigure[]{
		\hspace{-0.5cm}\epsfig{file=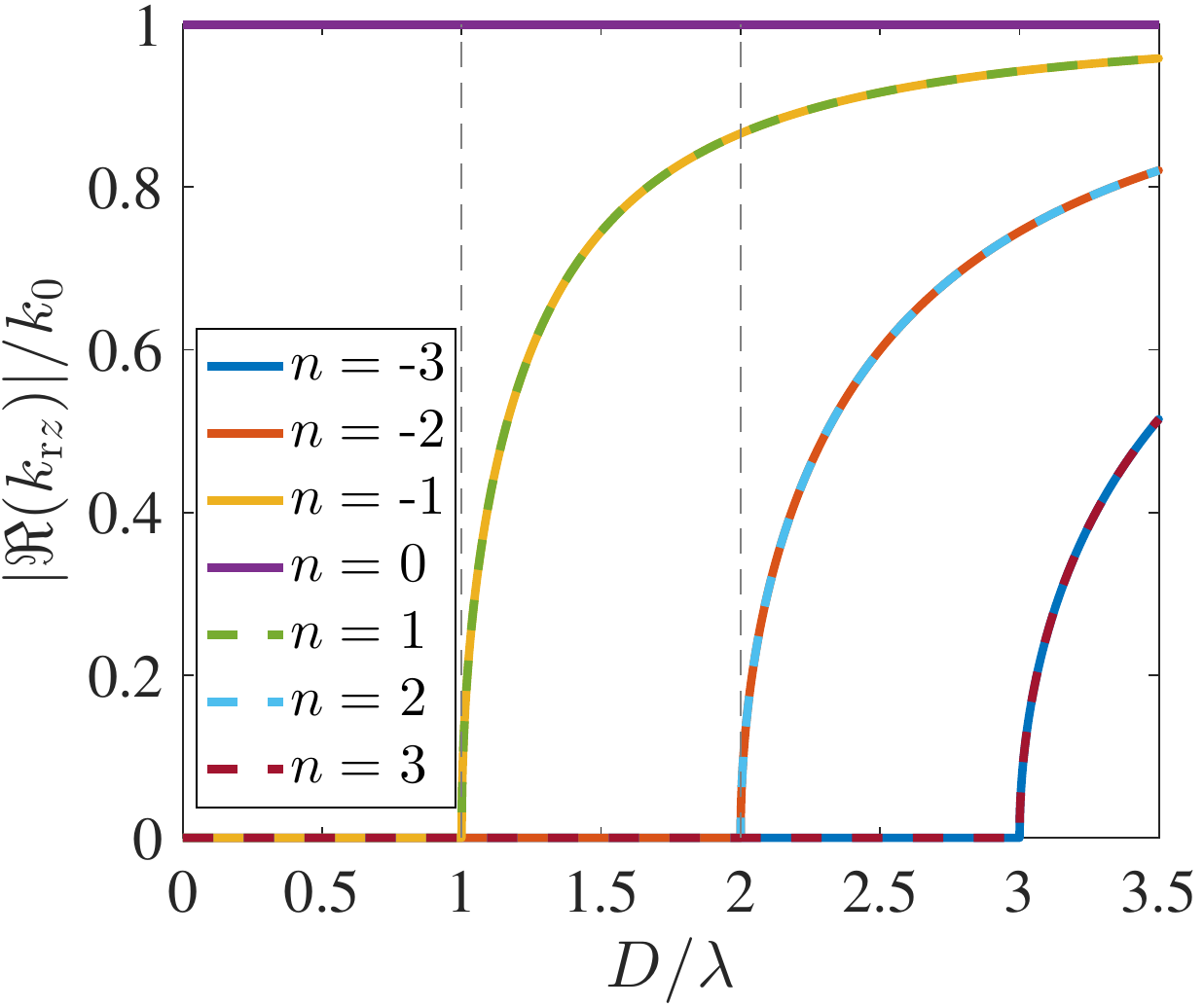, width=0.49\columnwidth}  
		\label{fig2a} }
	\subfigure[]{
		\epsfig{file=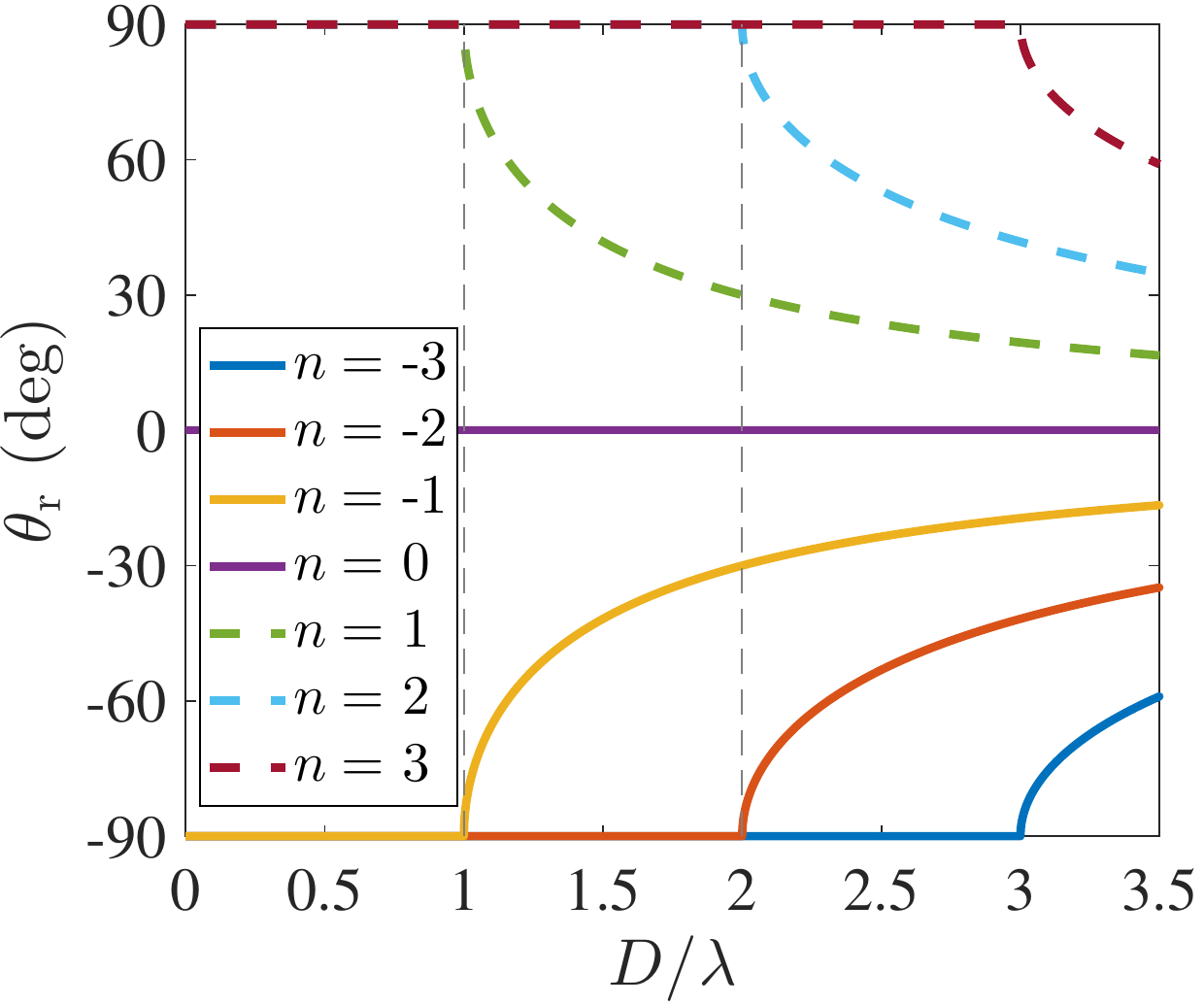, width=0.49\columnwidth} 
		\label{fig2b} }  
         \caption{(a) The real part of the normal wavenumber for different Floquet harmonics versus the periodicity of the reflector. (b) The reflection angle of the corresponding Floquet harmonics. The incidence angle is $\theta_{\rm i}=0^\circ$. 
	}\label{fig2}
\end{figure} 
The operation of the reflector strongly depends on the periodicity. To highlight it, we divide  the plots in Fig.~\ref{fig2} into three characteristic regions using vertical dashed lines. The first region  corresponds to reflectors with the periodicity smaller than the wavelength $\lambda$. Such reflectors (e.g., most natural materials, uniform antenna arrays and metasurfaces) illuminated normally exhibit  only the usual mirror reflection (harmonic $n=0$). The second region corresponds to flat reflectors with the period $\lambda<D<2\lambda$. Under normal illumination, they may provide anomalous reflection into the $+\theta_{\rm r}$ or  $-\theta_{\rm r}$ directions (harmonics $n=1$ and $n=-1$, respectively). Hereafter, we adopt the usual convention of counting $\theta_{\rm i}$ counter-clockwise and $\theta_{\rm r}$  clockwise from the $z$-axis (as  shown in Fig.~\ref{fig1}).   Most of the recently proposed gradient metasurface reflectors \cite{capasso,sun,bozh1,bozh2,mosall2,nanotechnology,prx4,veysi,
functional,li,synthesis,Alu,last,ana} operate in this periodicity region. Reflectors  with the periodicity $D>2\lambda$ (the third region)  illuminated normally have more than three ``open'' channels for propagating reflected waves. Reflectors operating in this region were not widely studied mainly due to the more complicated  design procedures (more channels where uncontrolled reflections  occur become ``open''). 
It should be noted that in the case of oblique incidence the number of propagating channels is generally different from that found for the normal incidence. For example, as it will be shown below, for a sub-wavelength periodicity $0.5 \lambda<D<\lambda$ more than one mode become  propagating.

\subsection{Fundamental classes of periodical reflectors}
Here we explore to what extent 
and using what metasurface topologies  
it is possible to engineer reflections through all open channels, which are defined by the surface modulation period. 
As an example, let us first concentrate on ``three-channel'' reflectors because of their simple and concise analysis (reflectors with the higher number of channels are discussed in Section~\ref{5port}). The three-channel regime can be realized in metasurfaces with a period $D<2\lambda$ if one of the channels corresponds to the normal incidence [see an illustration in  
Fig.~\ref{fig1b}].
It is convenient to represent the three propagation channels of this system as a three-port network using an analogy from the circuit theory. The channels are numbered according to Fig.~\ref{fig1b}. One can associate a scattering matrix with this system $S_{i j}$ ($i,j=1,2,3$) which measures the ratio of the \textit{tangential} electric field components (normalized by the square root of the corresponding impedance $Z_{i,j}$) of   the wave reflected into the $i$-th channel when the reflector is illuminated from  the $j$-th channel \cite{smatrix}, i.e., $S_{ij}=(E_{t,i}/\sqrt{Z_i})/(E_{t,j}/\sqrt{Z_j}) = \sqrt{E_{t,i}H_{t,i}}/\sqrt{E_{t,j}H_{t,j}}$.
 Assuming that the reflector is reciprocal and lossless, we have two constraints on the scattering matrix: It must be equal to its transpose $S_{i j}=S_{j i}$ and   unitary  $S_{k i}^* \,S_{k j} = \delta_{i j}$, where $\delta_{i j}$ is the Kronecker delta, and the  index notations are used. 
 
\subsubsection{General specular reflectors}
Obviously, for a usual mirror made of perfect electric conductor (PEC), the scattering matrix [with ports defined as in Fig.~\ref{fig1b}] is anti-diagonal with all non-zero elements equal to  the reflection coefficient of PEC, i.e.,  $S_{13}=S_{22}=S_{31}=-1$. Assuming possibly arbitrary reflection phases, in the most general form this matrix can be written as
\begin{equation}\label{m1} 
S = \begin{pmatrix} 
0 & 0 & e^{j \phi_{{\rm or}2}} \\ 0 & e^{j \phi_{{\rm or}1}} & 0 \\
e^{j \phi_{{\rm or}2}} & 0 & 0
     \end{pmatrix}.
\end{equation}
Here $\phi_{{\rm or}1}$ and $\phi_{{\rm or}2}$ represent phases of ordinary reflections when the reflector is illuminated normally and obliquely at $\theta_{\rm r}$, respectively. The matrix satisfies both previously defined constraints. In contrast to usual metal mirrors, the reflection phase in gradient reflectors described by (\ref{m1}) can be engineered arbitrarily and \emph{independently} for normal and oblique illuminations. In other words, there are no physical limitations that would forbid us from creating a reflector which, for example, operates as a metal mirror when illuminated normally and as a ``magnetic'' mirror \cite{magneticmirror} when illuminated obliquely. 

\subsubsection{Anomalous reflectors}
In the case of flat reflectors exhibiting ideal anomalous reflection \cite{ana}, the scattering matrix  can be fully determined using the symmetry constraints. Indeed, assuming that a normally illuminated reflector sends all the incident power to the $n=1$ channel and, reciprocally, power from  $n=1$ channel to the normal direction, we immediately fix its response for illumination from  $n=-1$ channel. 
The scattering matrix in this scenario reads
\begin{equation}\label{m2} 
S = \begin{pmatrix} 
0 & e^{j \phi_{\rm an1}} & 0  \\ 
e^{j \phi_{\rm an1}} & 0& 0 \\
0 & 0 & e^{j \phi_{\rm is1}}
     \end{pmatrix},
\end{equation}
where $\phi_{\rm an1}$ and $\phi_{\rm is1}$ are the phases of anomalous reflection and reflection for illumination from the $n=-1$ channel, respectively.

As is seen from~(\ref{m2}), the   $n=-1$ channel (port 3) is completely isolated from the other open channels of this metasurface. Incident light from the $-\theta_{\rm r}$ direction is always fully reflected back at the same angle. Moreover, no light from other directions can be reflected into this direction. This behaviour is  imposed by the reciprocity and energy conservation. 
As an example, we consider the anomalous reflector proposed in \cite{ana} which under normal illumination reflects 100\% of power at $\theta_{\rm r}=70^\circ$.  
Figure~\ref{fig3a} depicts the reflection angles for different propagating harmonics in this system versus the illumination angle $\theta_{\rm i}$. 
When the reflector is illuminated from port~1 or port~2, it creates an additional tangential momentum added to the incident wavevector $k_x=2\pi n /D$, where $n=+1$. 
The grey dashed line denotes the specular reflection angles for an equivalent mirror tilted at $\theta_{\rm i}/2=35^\circ$ (for such a mirror $\theta_{\rm r}=\theta_{\rm i}+70^\circ$). As is expected, the reflector imitates the behaviour of a titled mirror at two angles, when $\theta_{\rm i}$ equals to $0^\circ$ and $-70^\circ$ (marked by circles in the plot). Figure~\ref{fig3b} illustrates this result by eyes of an external observer. 
\begin{figure}[h]
	\centering
	\subfigure[]{
		\epsfig{file=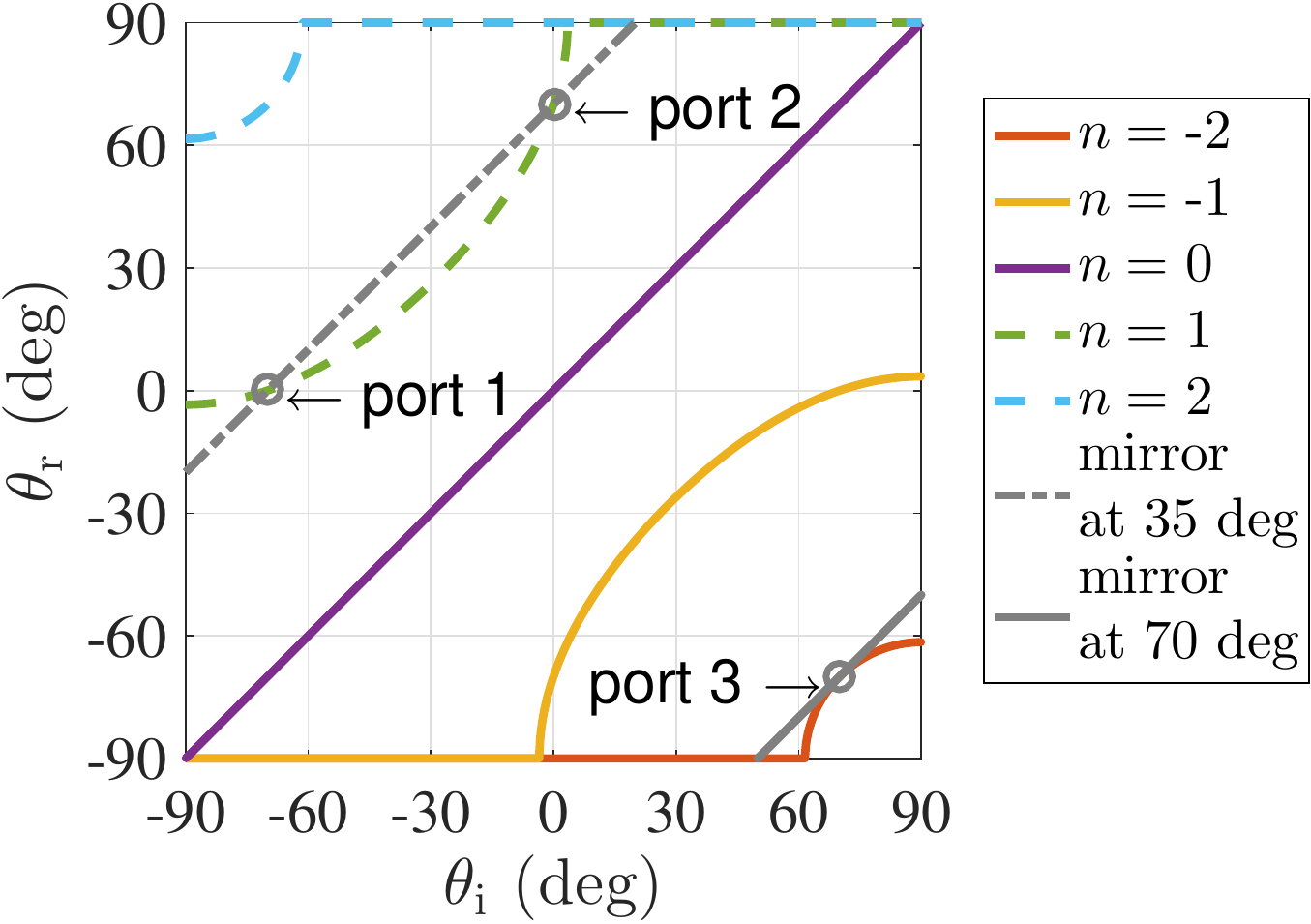, width=0.49\columnwidth}  
		\label{fig3a} }
	\subfigure[]{
		\epsfig{file=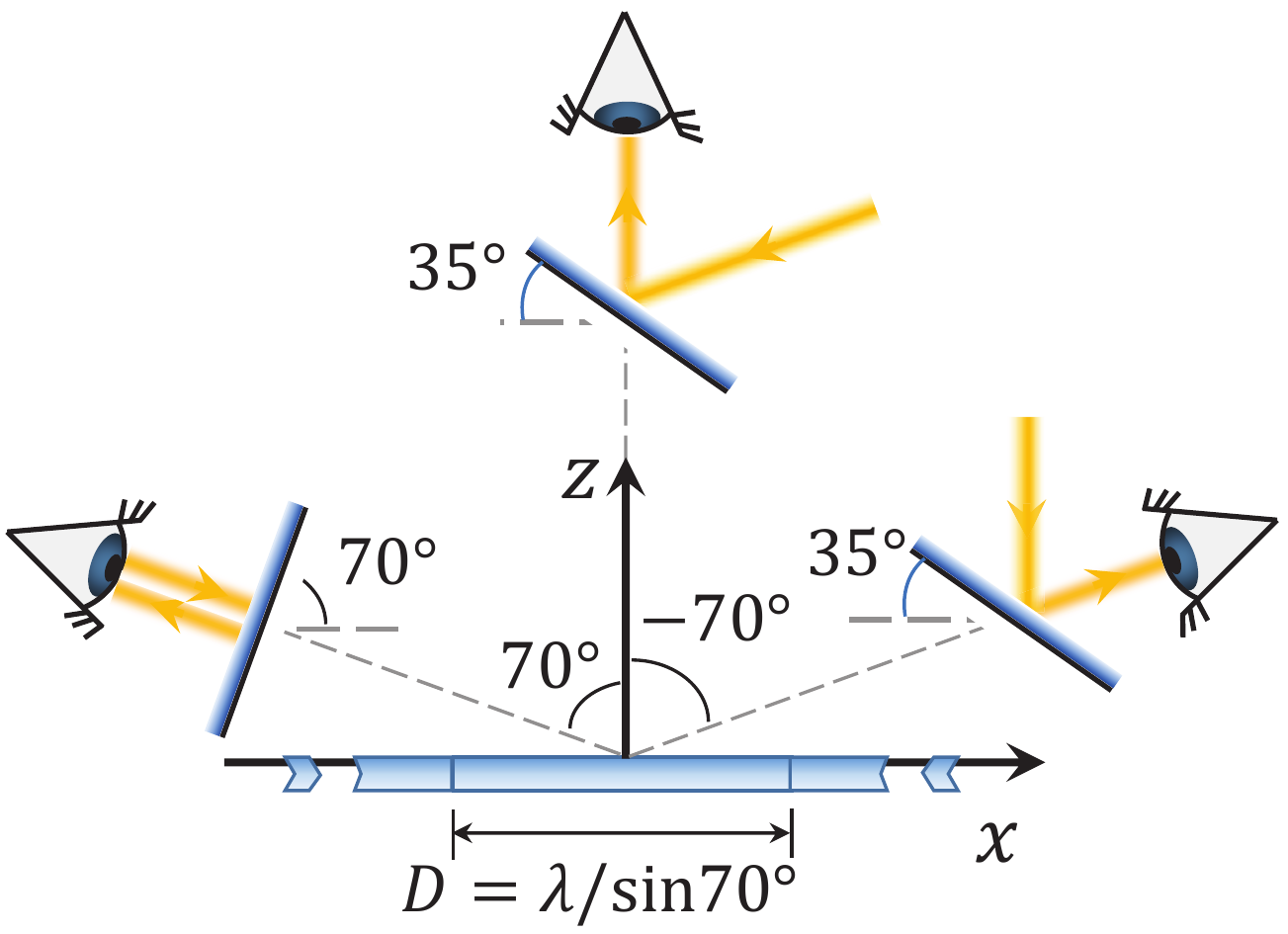, width=0.43\columnwidth} 
		\label{fig3b} }  
         \caption{(a) Operation of the anomalous reflector proposed in \cite{ana}. Reflection angle versus incidence angle for different Floquet harmonics. All the harmonics are considered. The circles show at what angle $\theta_{\rm r}$ \textit{all} the incident energy is reflected from the considered anomalous reflector when it is excited from the corresponding port. (b) The flat anomalous reflector appears to an external observer at different angles as differently tilted mirrors. At $\theta_{\rm i}=0^\circ$ and $\theta_{\rm i}=-70^\circ$ angles, it appears as a tilted at $35^\circ$ mirror. However, at $\theta_{\rm i}=70^\circ$ angle, the observer would see himself as in a mirror tilted at $70^\circ$.
	}\label{fig3}
\end{figure} 
However, when $\theta_{\rm i}=+70^\circ$ (from port 3),  the reflector emulates a mirror titled at $70^\circ$ (for such a mirror $\theta_{\rm r}=\theta_{\rm i}-140^\circ$) which corresponds to the  solid grey line in Fig.~\ref{fig3a}. 
In this case the $n=+1$ mode is evanescent ($k_{{\rm r} x}> k_{\rm i}$), while the $n=-2$ mode becomes propagating and the reflector creates a negative tangential momentum $k_x=-4\pi  /D$ responsible for retroreflection. 
Thus, an external observer looking at the reflector at $\theta_{\rm i}=70^\circ$ would see his/her own image [see Fig.~\ref{fig3b}].
 Experimental verification of this isolation property of the reflector proposed in \cite{ana} is demonstrated in supplementary materials \cite{suppl}.  At all other angles, the reflection from the structure constitutes a combination of  several harmonics whose amplitudes depend on the specific design of the reflector. Interestingly, the isolation effect of anomalous reflectors was not widely discussed in literature. A recent study \cite{sun} erroneously reports that when such a reflector is illuminated from port~3, it forms a surface wave bound to the reflector.

Likewise, the normally incident power can be   redirected to the  $n=-1$ channel, yielding the scattering matrix
\begin{equation}\label{m21} 
     S = \begin{pmatrix} 
e^{j \phi_{\rm is2}} & 0 & 0  \\ 
0 & 0& e^{j \phi_{\rm an2}} \\
0 & e^{j \phi_{\rm an2}} & 0
     \end{pmatrix},
\end{equation}
where $\phi_{\rm an2}$ and $\phi_{\rm is2}$ are the phases of anomalous reflection and reflection for illumination from the  $n=1$ channel, respectively. It should be noted that scattering matrices (\ref{m2}) and (\ref{m21}) represent equivalent physical properties. Matrix (\ref{m21}) corresponds to the metasurface modelled by (\ref{m2}) but rotated by $180^\circ$ around the $z$-axis.

\subsubsection{Three-channel retro-reflectors}
By analysing the structure of scattering matrices (\ref{m1}), (\ref{m2}) and (\ref{m21}), we can observe that one more matrix form with three non-zero components is possible, which satisfies the symmetry constraints of the reciprocity and energy conservation:
\begin{equation}\label{m3} 
S = \begin{pmatrix} 
e^{j \phi_{\rm is 1}} & 0 & 0  \\ 
0 & e^{j \phi_{\rm is 2}}& 0 \\
0 & 0 & e^{j \phi_{\rm is 3}}
     \end{pmatrix},
\end{equation}
where $e^{j \phi_{\rm is 1,2,3}}$ denote independent reflection phases for different illumination angles. 
It is simple to check that all other tensor structures with three non-zero components are forbidden. Remarkably,  the three classes of   reflectors with scattering matrices defined by (\ref{m1}), (\ref{m2}) and (\ref{m3}) [matrix (\ref{m21}) is equivalent to (\ref{m2})] constitute a peculiar basis of most general three-channel reflectors. 
 From the physical point of view, the three \emph{elemental} functionalities of these reflectors form a basis of all possible reflection functionalities achievable with such periodic flat structures. 
Moreover, it can be shown that the response of an arbitrary multi-channel ($N \times N$) reflector can be described by a combination of the three  basic functionalities: General specular, anomalous, and retro-reflection. 


\vspace{0.5cm}
\section{Three-channel retro-reflector} \label{three}
A closer look at scattering matrix (\ref{m3}) of a three-channel retro-reflector reveals that it corresponds to a so-called ``isolating'' mirror: In contrast to the previously considered structures, here all three channels are isolated from one another. This retro-reflector (or isolating mirror) under the assumption of lossless response represents a sub-wavelength blazed grating in the Littrow mount \cite{grating1,grating2,retro1}. Conventional realizations of Littrow  gratings employ triangular or binary groove profiles with the total thickness of several wavelengths (metallic or dielectric gratings)~\cite{retro1,retro2,retro3}. While this thickness is practically acceptable for optical gratings, it becomes very critical for their microwave counterparts resulting in several-centimeters-thick structures. Therefore, for microwave and radio frequency applications there is a need in sub-wavelength thin designs. A good candidate for low-frequency multi-channel mirrors are flat reflectarray antennas of slim profile~\cite{encinar}. 
However, to the best of our knowledge, there is only one work devoted to the design of flat reflectarrays comprising a dense array of sub-wavelength resonators  redirecting energy in the direction of incidence \cite{fromana}, and    the isolating nature of the system was  not explored in this work. 
In comparison, traditional retro-directive array antennas typically use a half-wavelength element spacing. In addition, they employ a collection of delay lines in passive designs or an active phase-conjugating circuitry in active designs behind the array plane~\cite{Miyamoto}.
Therefore,  we next rigorously design a flat three-channel retro-reflector (isolating mirror) and experimentally investigate its response from all  three open-channel directions. 

Here,    the general approach \cite{synthesis} based on the surface impedance concept is used. 
The design methodology starts with the definition of the total fields at the metasurface plane that ensure the desired functionality of port~3  (alternatively, it could be port~1). We require that \textit{no evanescent waves} are excited for illumination from port~3 and the reflection constitutes a single plane wave. 
Considering as an example the TE polarization (electric field polarized along the $y$-axis), one can relate the tangential total electric and magnetic fields through the surface impedance $Z_{\rm s}$:
\begin{equation}
\begin{array}{l}
\displaystyle
E_{\rm i}e^{-jk_{\rm i} \sin \theta_{\rm i}x}+E_{\rm r}e^{jk_{\rm i} \sin \theta_{\rm i}x}\vspace*{.2cm}\\\displaystyle
\hspace*{2cm}
=Z_{\rm s}\frac{\cos\theta_{\rm i}}{\eta}\left(E_{\rm i}e^{-jk_{\rm i} \sin \theta_{\rm i}x}-E_{\rm r}e^{jk_{\rm i} \sin \theta_{\rm i}x}\right),
\end{array}\label{eq:4}
\end{equation}
where $E_{\rm i}$ and $E_{\rm r}$ are the amplitudes of the incident and reflected waves at port~3, 
and $\eta$ is the intrinsic impedance in the background medium. Ensuring that all the input energy is reflected back into the same direction $E_{\rm i}=E_{\rm r}$,  the surface impedance reads $Z_{\rm s}=j\frac{\eta}{\cos\theta_{\rm i}}\cot(k_{\rm i} \sin \theta_{\rm i}x)$. Such  surface impedance can be  realized as a 
single-layer structure. According to the theory of high-impedance surfaces \cite{Sievenpiper}, an arbitrary reactive surface impedance can be achieved using a capacitive metal pattern separated from a metal plane by a thin dielectric layer.
Since the cotangent function repeats with a period $\pi$, the  periodicity of this impedance profile is $D=\lambda/(2\sin\theta_{\rm i})$. This analytical solution gives periodicity  smaller than the wavelength if port~3 is at an angle $\theta_{\rm i}>30^\circ$ from the normal. 
It should be noted that the derived impedance expression is somewhat similar to that for conventional reflectarrays  obtained in \cite{ana}. This similarity appears because  both formulas were derived under the assumption that the metasurface possesses local electromagnetic response. 

For the  actual implementation, we design a three-port retro-reflector whose ports 1, 2, and 3 are directed at angles $-70^\circ$,  $0^\circ$, and  $+70^\circ$, respectively [see Fig.~\ref{fig1b}]. Due to the sub-wavelength periodicity $D=\lambda/(2\sin 70^\circ)$ imposed on the system, port~2  is isolated from the other two ports. The functionality of port~1, in the absence of dissipation loss, is automatically satisfied because of the reciprocity of the system.  
The same scenario occurs in lossless blazed gratings in the Littrow mount \cite{grating1,grating2,retro1}. Interestingly, if the metasurface is lossy, this conclusion is not generally correct and reflections in port~1 and port~2 in principle could be  tuned differently and independently. 
Note that there are other possible solutions for  lossless isolating mirrors with periodicity $\lambda<D<2\lambda$, however, they imply  excitation of evanescent waves when the  illumination is from either of the three channels, which complicates the theoretical analysis.

Figure~\ref{fig4} shows numerical simulations \cite{HFSS} of a  retro-reflector modelled as an ideal inhomogeneous sheet with the surface impedance calculated from Eq.~\ref{eq:4}. Figures~\ref{fig4}(a--c) depict the   real part of the scattered electric field (the instantaneous value), while Figs.~\ref{fig4}(d--f) show 
the amplitude of this field.
 When the metasurface is illuminated from port~3 [see Figs.~\ref{fig4a} and \ref{fig4d}], a perfect reflected plane wave (no evanescent waves) is generated in the same direction, fulfilling the design condition. Figures~\ref{fig4b} and \ref{fig4c} show the scattered electric fields when the metasurface is illuminated at $\theta_{\rm i}=0^\circ$ and $\theta_{\rm i}=-70^\circ$, respectively.  Likewise, in these scenarios the incident wave is fully reflected back at the same angle. However, as it is seen from Figs. \ref{fig4e} and \ref{fig4f},  evanescent fields naturally appear in order to satisfy the boundary conditions at the metasurface.

\begin{figure}[h]
	\centering
	\subfigure[]{
		\epsfig{file=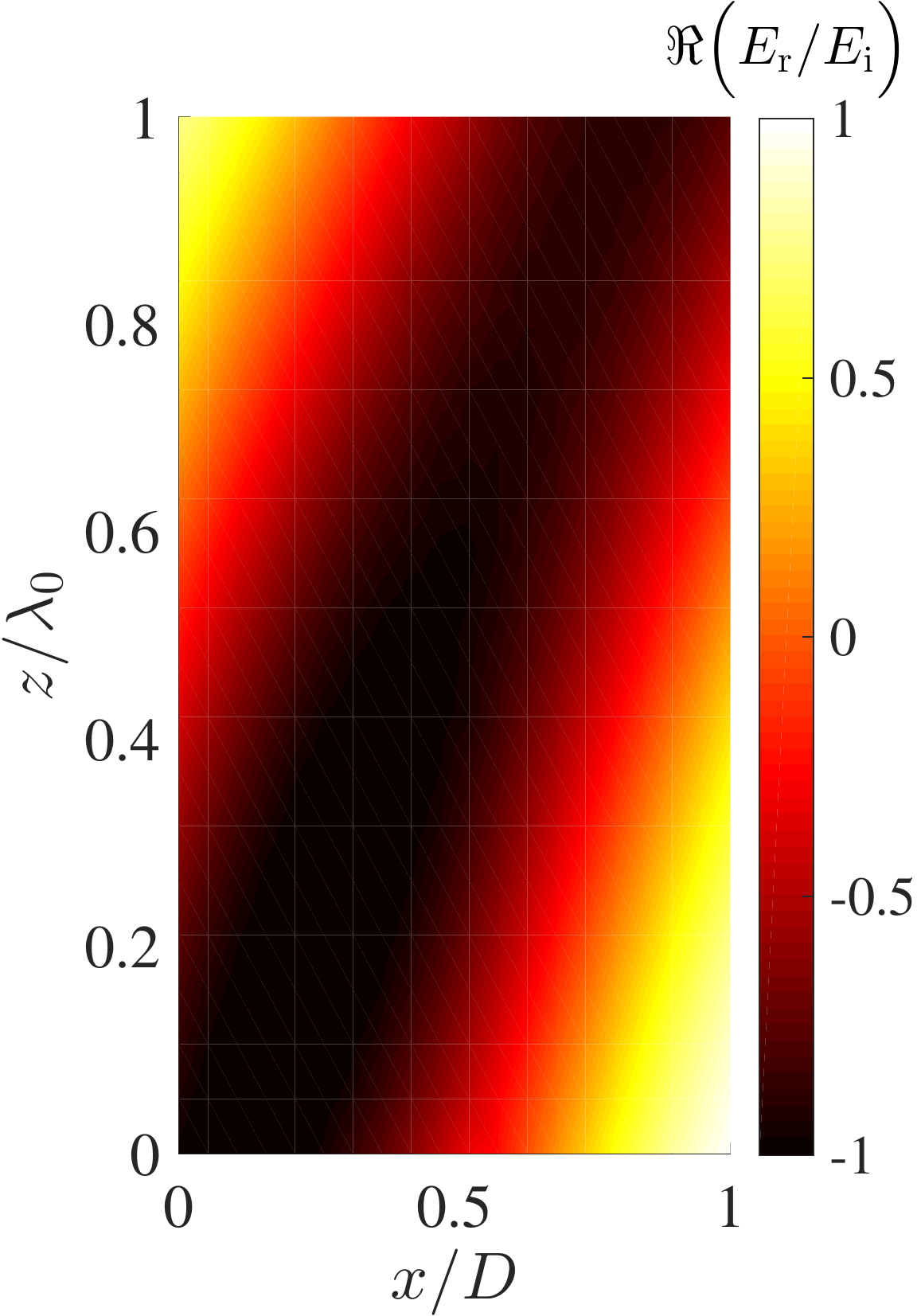, width=0.3\columnwidth}  
		\label{fig4a} }
	\subfigure[]{
		\epsfig{file=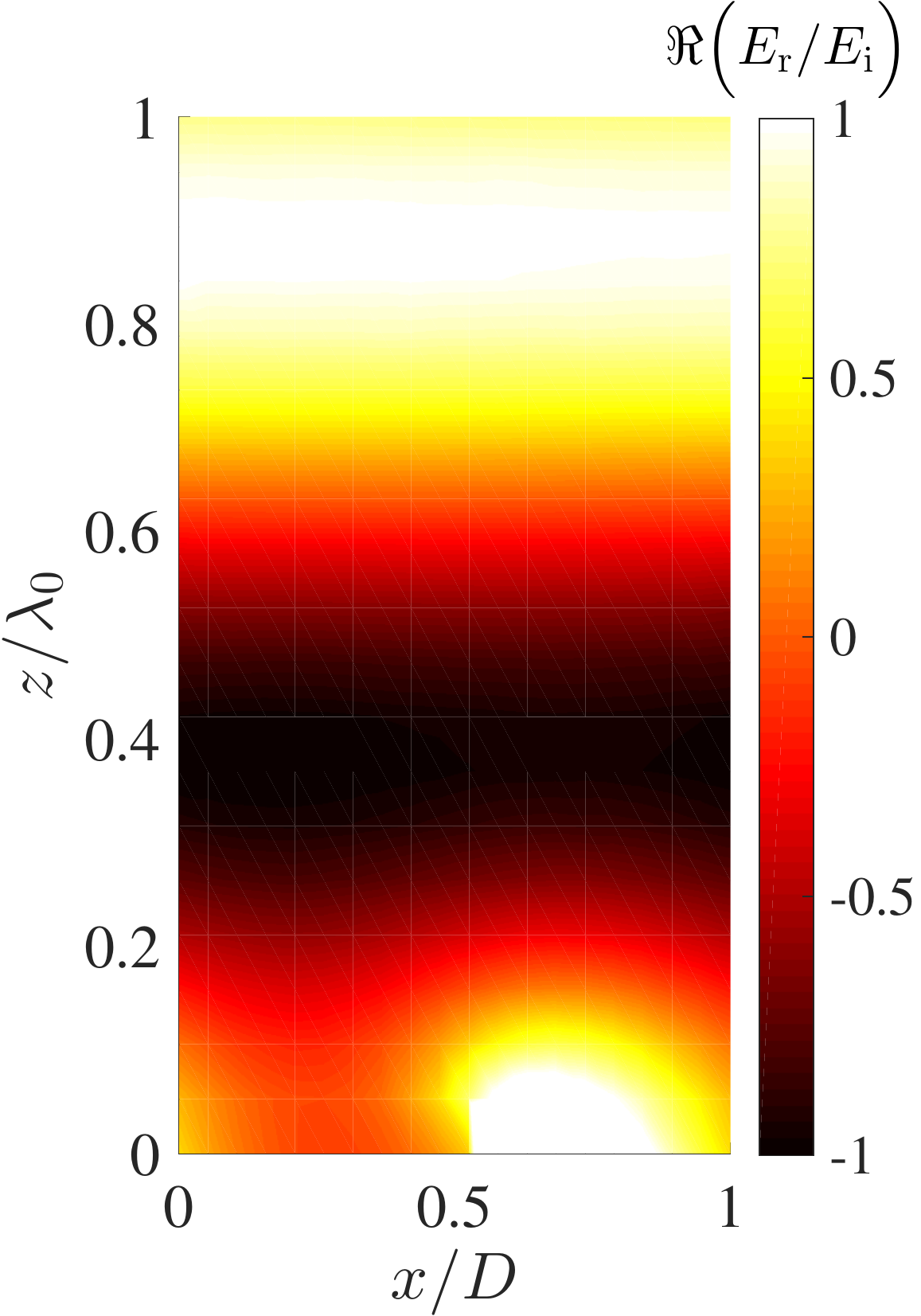, width=0.3\columnwidth} 
		\label{fig4b} }  
    \subfigure[]{
	    \epsfig{file=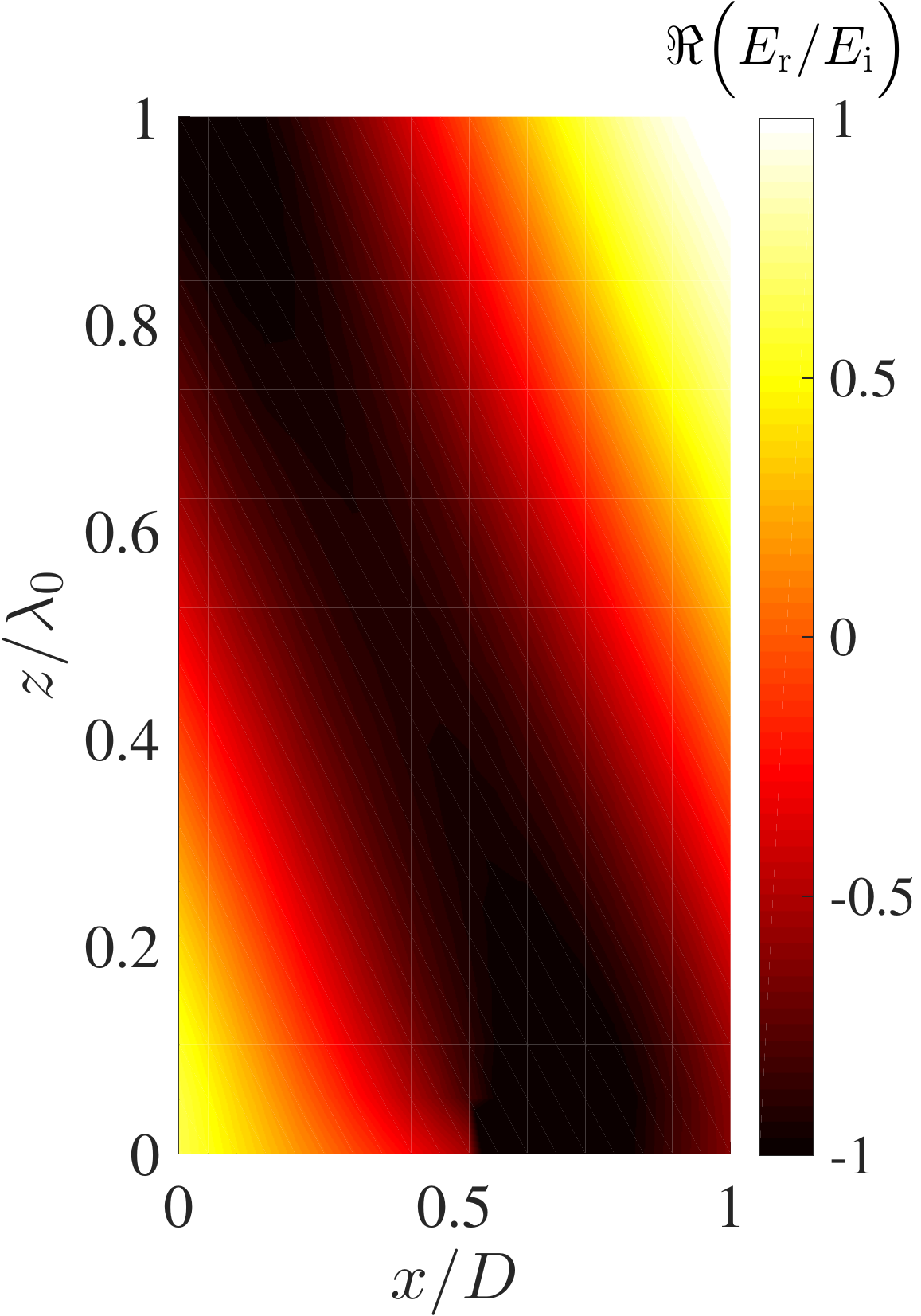, width=0.3\columnwidth} 
	\label{fig4c} } 
	\subfigure[]{
	\epsfig{file=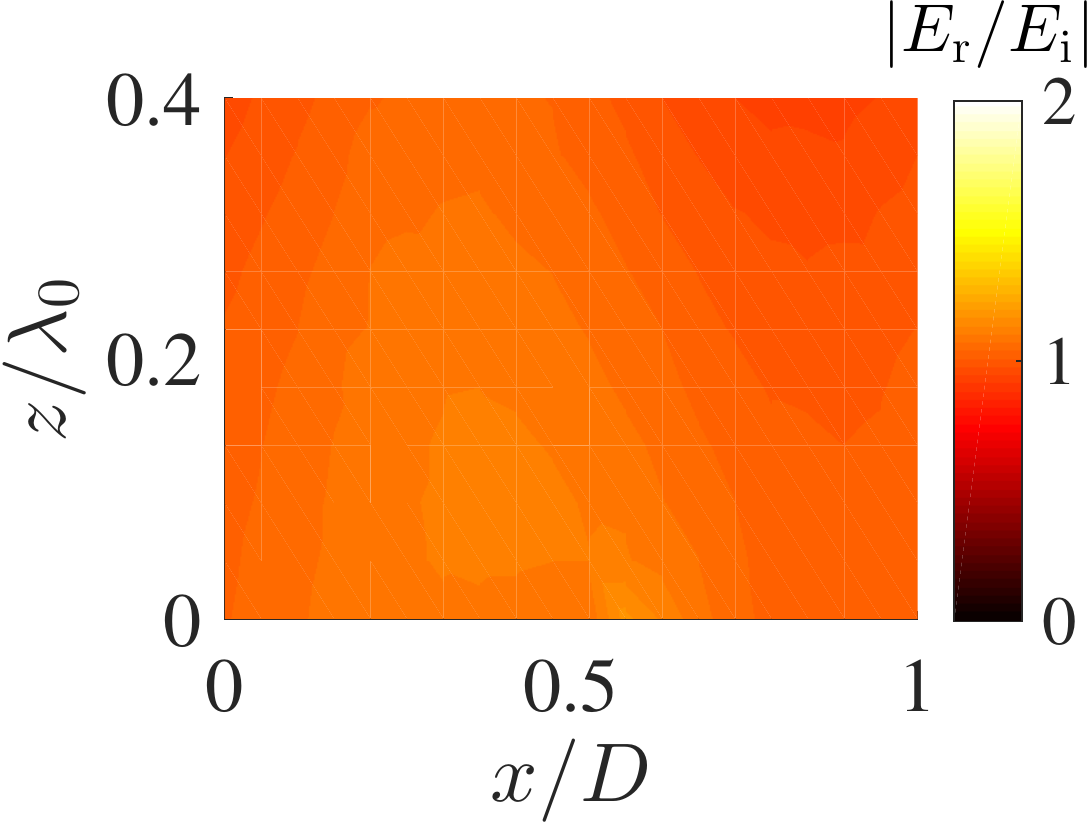, width=0.3\columnwidth}  
		\label{fig4d} }
	\subfigure[]{
	\epsfig{file=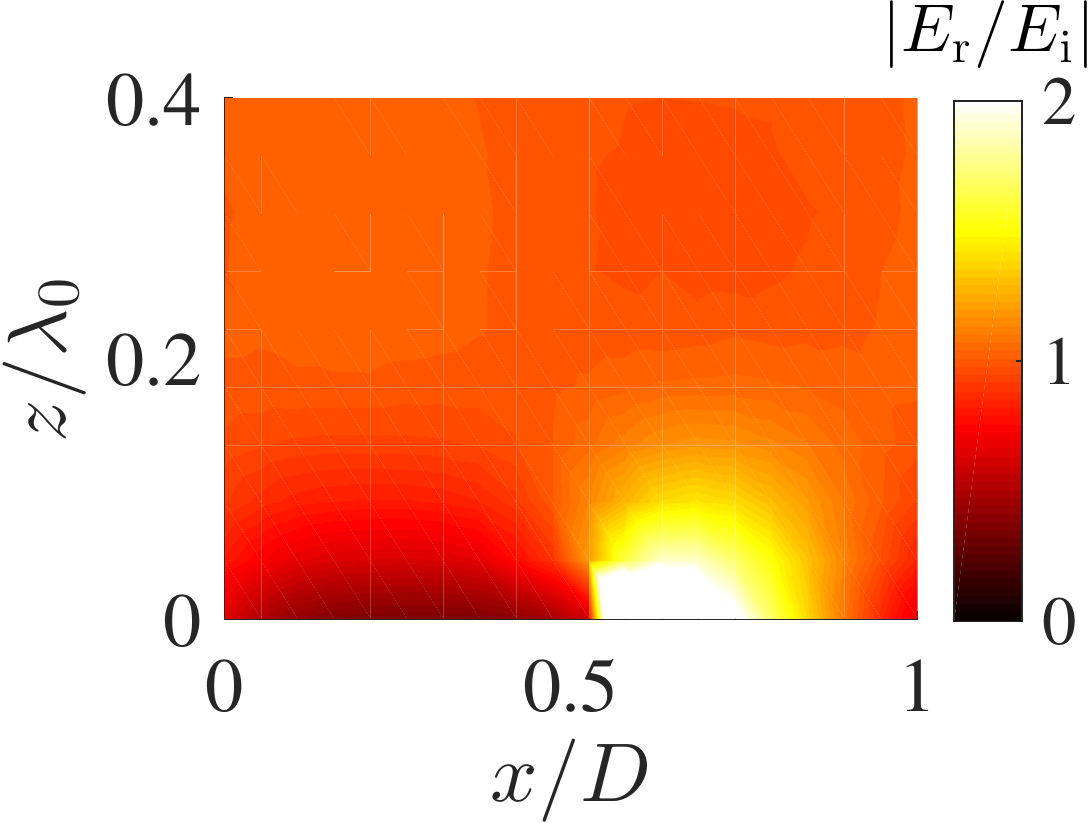, width=0.3\columnwidth} 
		\label{fig4e} }  
	\subfigure[]{
	\epsfig{file=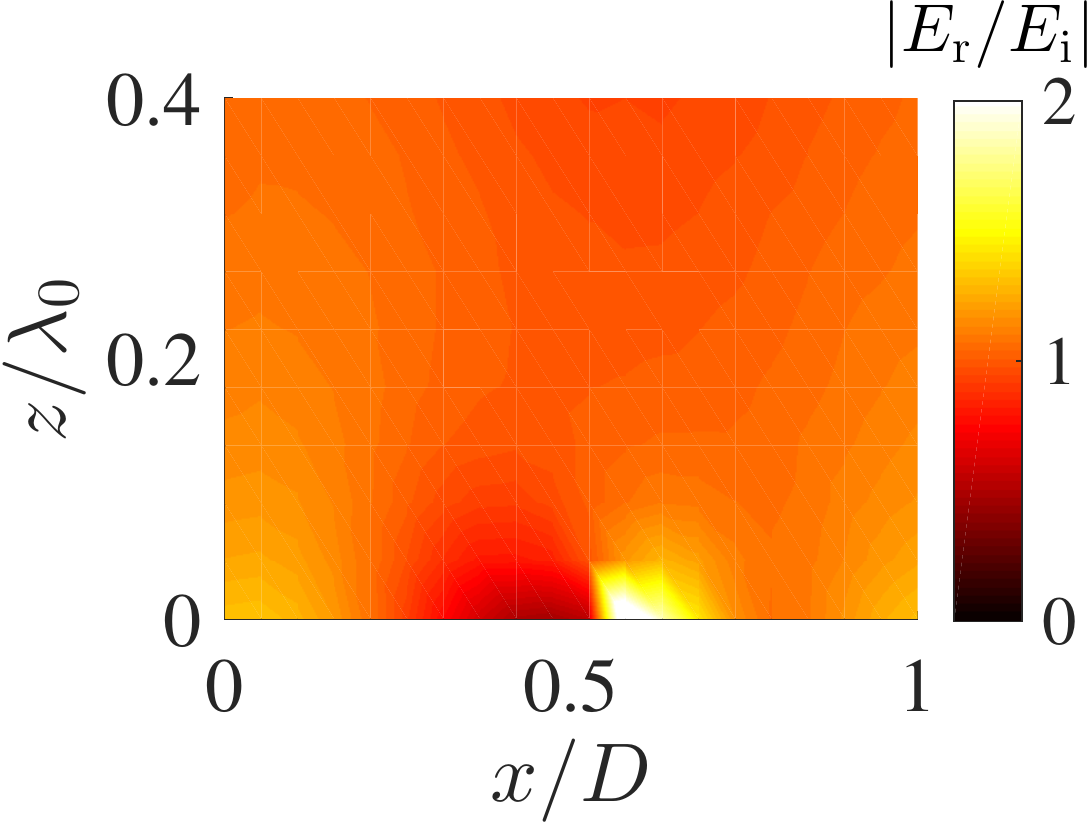, width=0.3\columnwidth} 
		\label{fig4f} }   
	\caption{Numerical simulations of the three-channel mirror modelled as an inhomogeneous sheet with surface impedance calculated from Eq.~(\ref{eq:4}). Real part (instantaneous value) and magnitude of the scattered normalized electric field when the reflector is illuminated from port 3 [(a) and (d)], port 2 [(b) and (e)] and port 1 [(c) and (f)].}\label{fig4}
\end{figure}

The purely imaginary surface impedance facilitates simple implementation of our design using conventional techniques. For demonstration purpose, we implement a three-port isolating mirror in the microwave regime ($8~{\rm GHz}$) using rectangular conducting patches over a metallic plane separated by a dielectric substrate [see Fig.~\ref{fig5a}]. 
\begin{figure}[h]
	\centering
	\subfigure[]{
		\epsfig{file=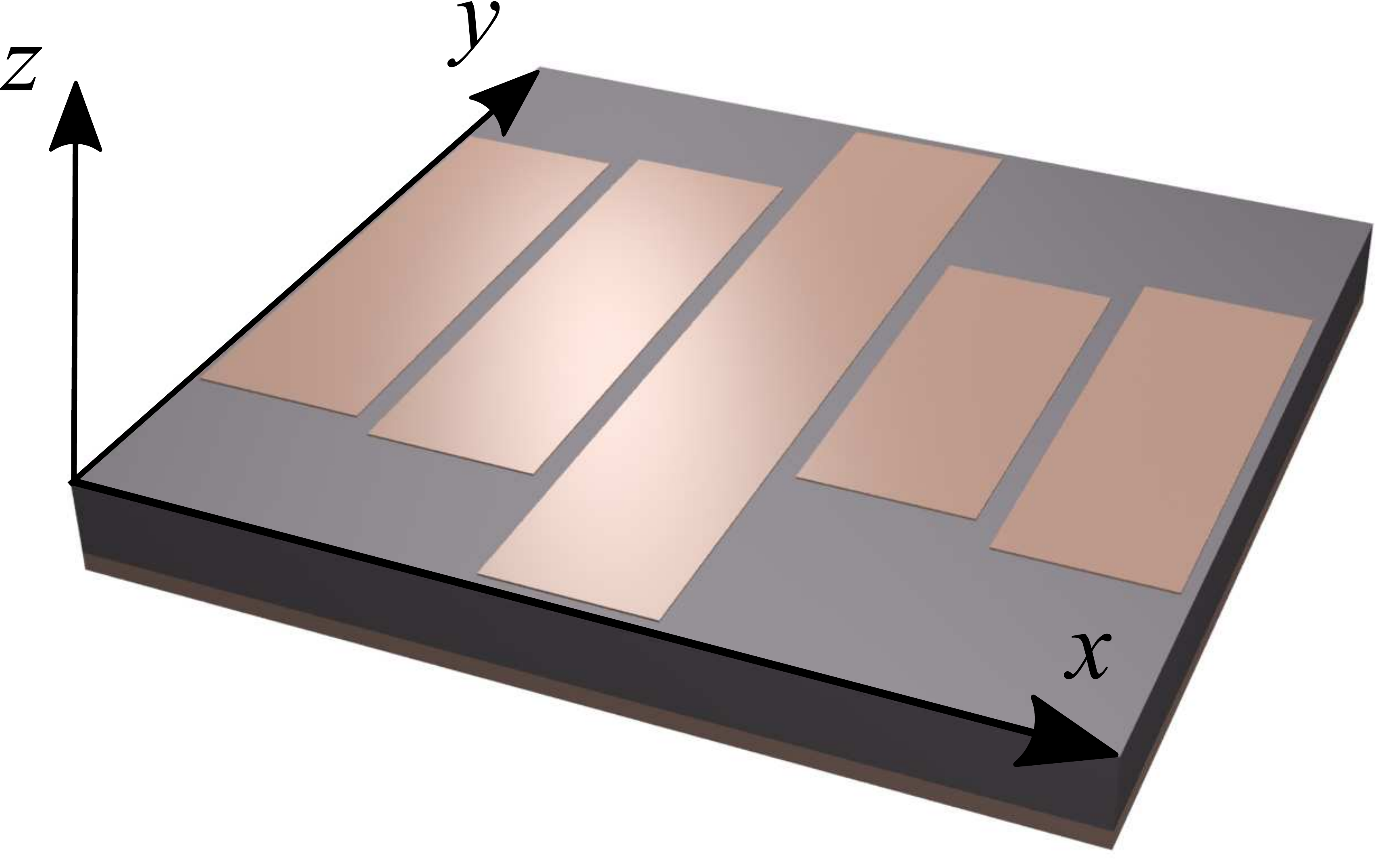, width=0.49\columnwidth}  
		\label{fig5a} }
	\subfigure[]{
		\epsfig{file=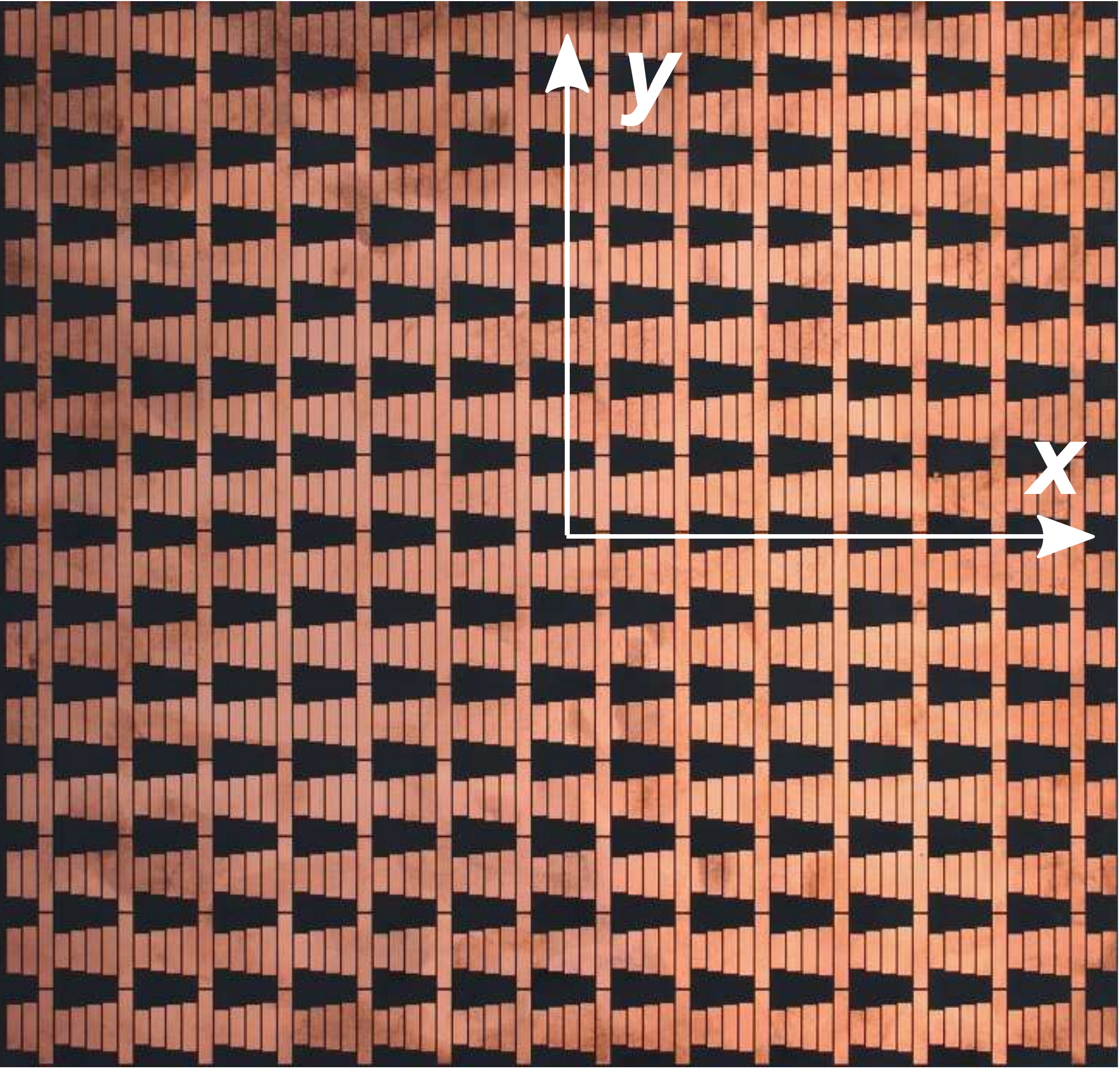, width=0.43\columnwidth} 
		\label{fig5b} }  
	\caption{(a) Schematic representation of one unit cell of the designed three-channel isolating mirror (retro-reflector). (b) A photograph of the fabricated prototype. }\label{fig5}
\end{figure}
Each unit cell consists of five patches with the same width of $3.5~{\rm mm}$  and different lengths aligned along the $y=0$ line. The dimensions of the unit cell  are $D=\lambda/(2\sin 70^\circ)=19.95$~mm and $\lambda/2=18.75$~mm along the $x$-axis and $y$-axis, respectively. For designing the array, each patch was placed in a homogeneous array of equivalent patches and the length was calculated to ensure the reflection phase dictated by the surface impedance of the system. Although the efficiency of the metasurface synthesized using this approach was high (about 90\% of retro-reflection in each channel), it was subsequently improved by numerical post-optimization of the patch dimensions. The numerically calculated efficiency of  retro-reflection of the final metasurface is 92.8\%, 95.4\%, and 94.3\% when excited from ports 1, 2, and 3, respectively. The rest of the energy is absorbed in the metasurface.
The final   lengths of the patches are  $11$, $11.8$, $18.1$, $8.4$, and $9.8~{\rm mm}$ (listed along the $x$-axis). The substrate material between the patches and the metallic plane is Rogers 5880 ($\varepsilon_{\rm d}=2.2$, $\tan \delta=0.0009$) with thickness $1.575~{\rm mm}$ ($\lambda/24$ at 8~GHz). Figure~\ref{fig6a} presents the results of numerical simulations of the final design. 
\begin{figure}[t]
	\centering
	\subfigure[]{
		\epsfig{file=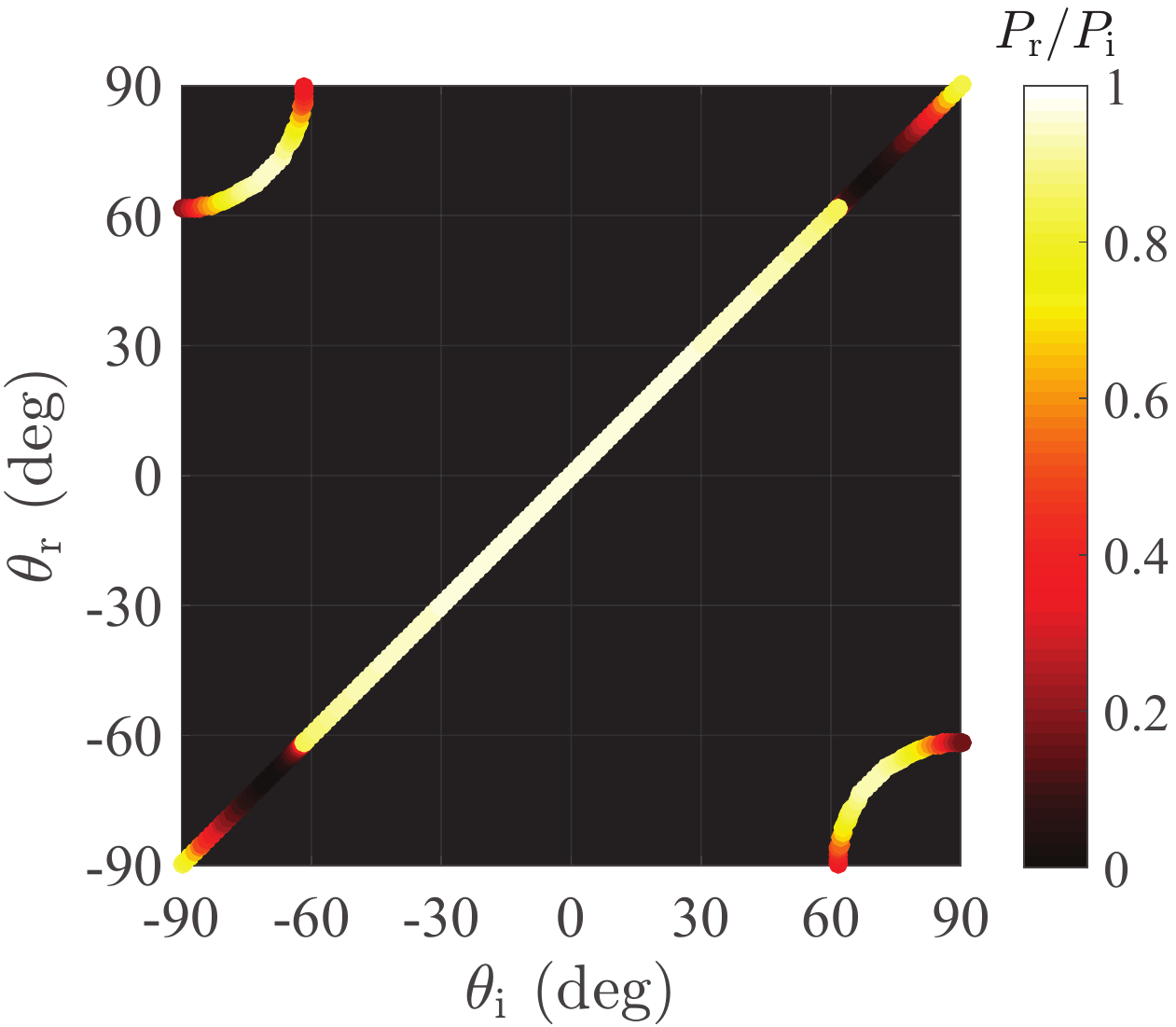, width=0.49\columnwidth}  
		\label{fig6a} }
	\subfigure[]{
		\epsfig{file=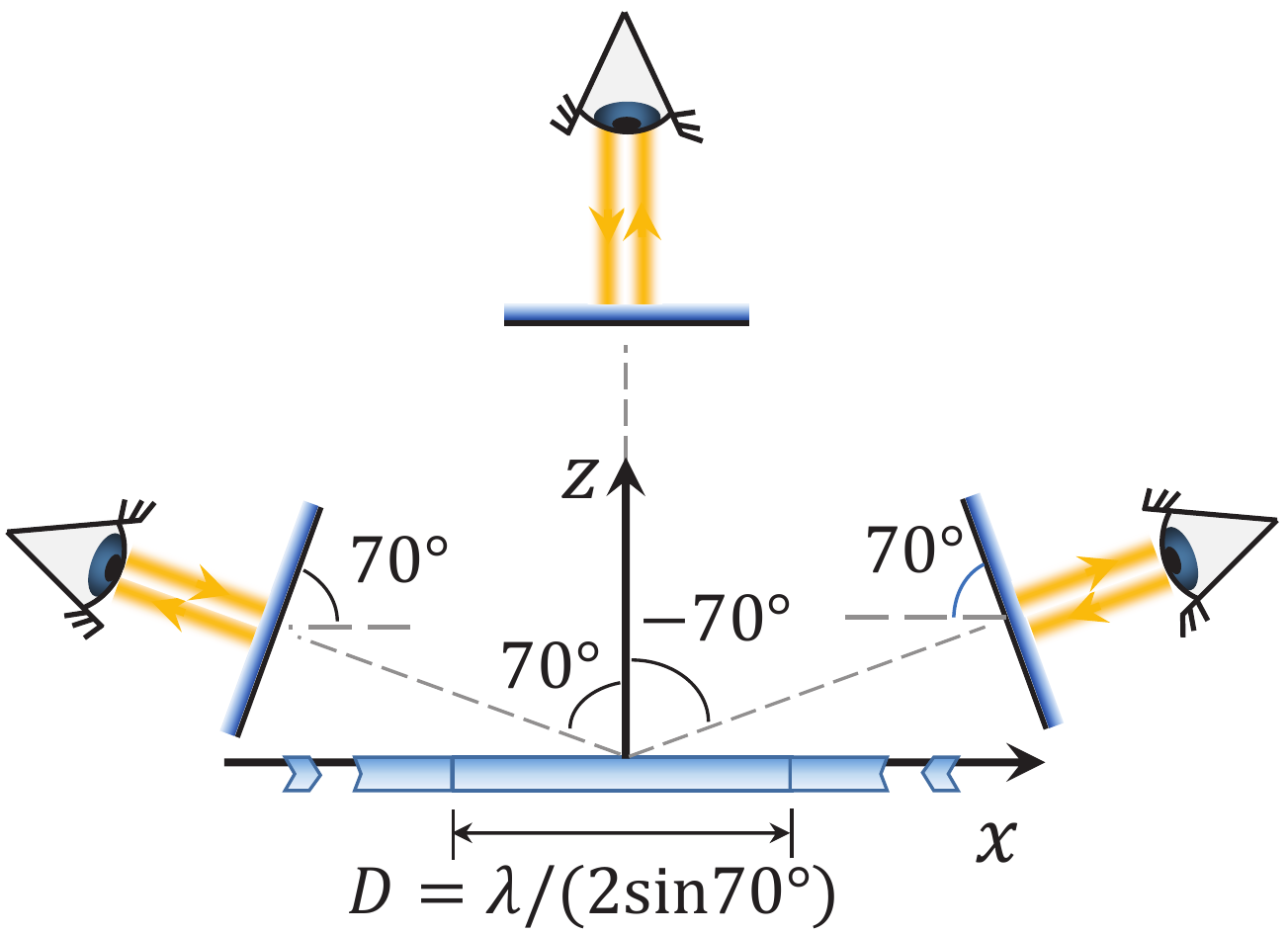, width=0.43\columnwidth} 
		\label{fig6b} }  
         \caption{(a) Simulated results of the three-channel retro-reflector made of a patch array. Distribution of the normalized reflected power across Floquet harmonics propagating at different angles $\theta_{\rm r}$ versus incidence angle $\theta_{\rm i}$. (b) Appearance of the flat isolating mirror for an external observer.  At $\theta_{\rm i}=0^\circ$ and $\theta_{\rm i}=\pm 70^\circ$ angles, the observer would see himself as in a mirror normally oriented with respect to him.
	}\label{fig6}
\end{figure}
The colour map represents the normalized reflected power into the $\theta_{\rm r}$ direction when the system is illuminated from the  $\theta_{\rm i}$ direction.  We can clearly identify the directions of the three open channels as the light coloured regions when $\theta_{\rm i}=-\theta_{\rm r}$.
From the colour map one can see that the metasurface reflects energy back in the diffracted (non-specular) direction  when illuminated not only at $\pm 70^\circ$, but in the range of angles from around $\pm 62^\circ$ to $\pm 90^\circ$ (the amount of this  energy decreases when $\theta_{\rm i}$ deviates from $\pm 70^\circ$). For example, when $\theta_{\rm i}=75^\circ$, about 88.8\% is diffracted back at the angle $\theta_{\rm r}=-66^\circ$. Such diffraction regime  resembles reflection from a metal mirror tilted at $70^\circ$ (to be precise, this mirror would have some curvature since $\theta_{\rm i} \neq -\theta_{\rm r}$). 
Extending the previously shown explanations,  Figure~\ref{fig6b} schematically represents the appearance of the flat three-channel isolating mirror for an external observer. For simplicity, the curvature of the equavalent mirrors tilted at $\pm 70^\circ$ is omitted in the illustration.
Looking along any of the three directions of open channels, the observer always will ``see" a  mirror orthogonal to the observation direction. 
Notice that in contrast to conventional blazed gratings whose design procedure  based on the coupled-mode theory is approximate and requires heavy post-optimization, our synthesis method provides a straightforward and rigorous solution (an exception is work~\cite{retro3} which appeared during the review process of the current paper).

The metasurface sample was manufactured using the conventional printed circuit board technology  and comprised $14\times14$  unit cells in the $xy$-plane [see Fig.~\ref{fig5b}] with the total dimensions of $7.5\lambda=282$~mm and $7\lambda=262.5$~mm along the $x$ and $y$ axes, respectively. 
The experiment was performed in an anechoic chamber emulating the free-space environment. The orientation of the sample, defined by the angle $\theta_{\rm i}$, was controlled by a  rotating platform  as shown in Fig.~\ref{fig7a}. 
\begin{figure}[t]
	\centering
	\subfigure[]{
		\epsfig{file=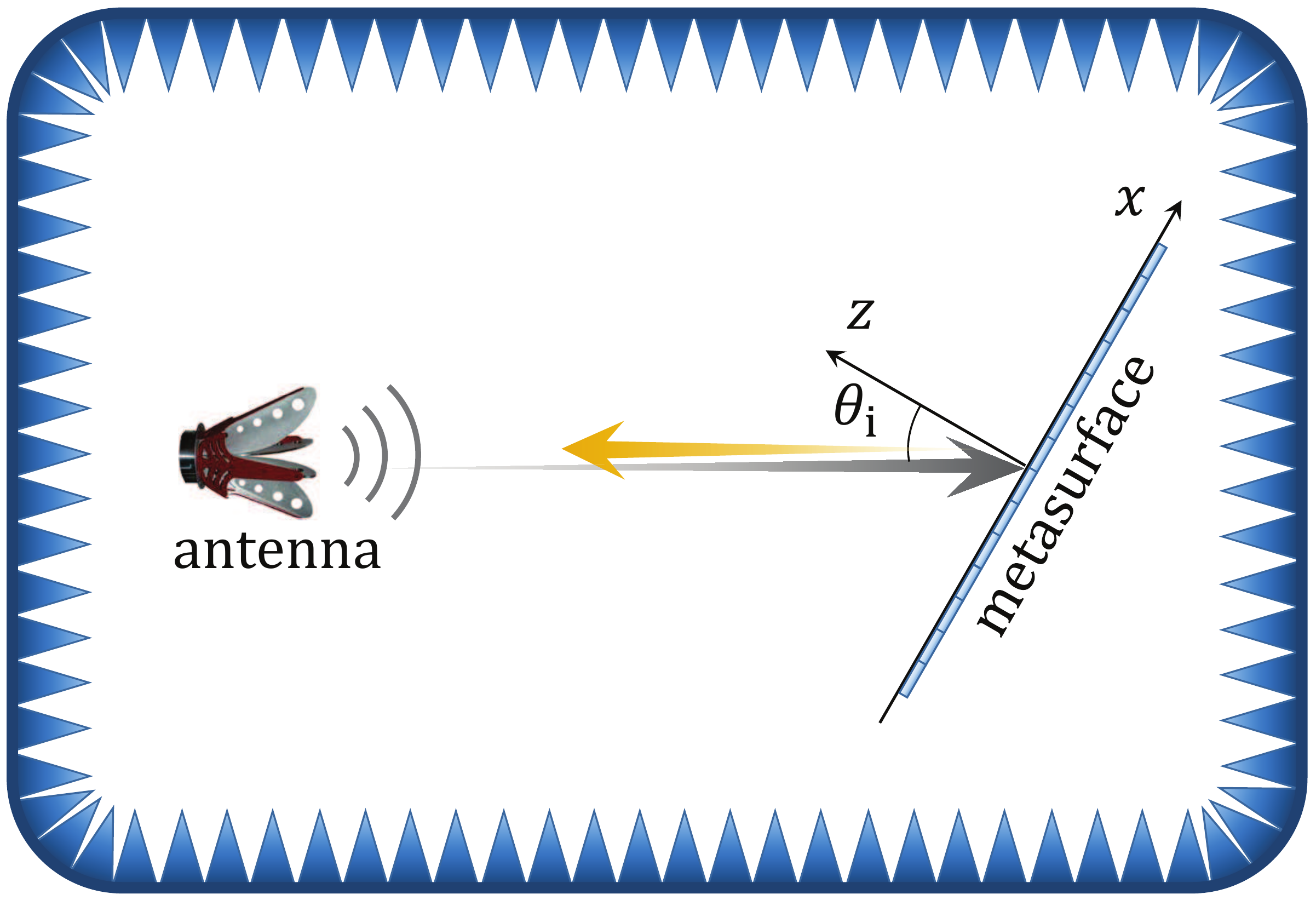, width=0.46\columnwidth}  
		\label{fig7a} }
	\subfigure[]{
		\epsfig{file=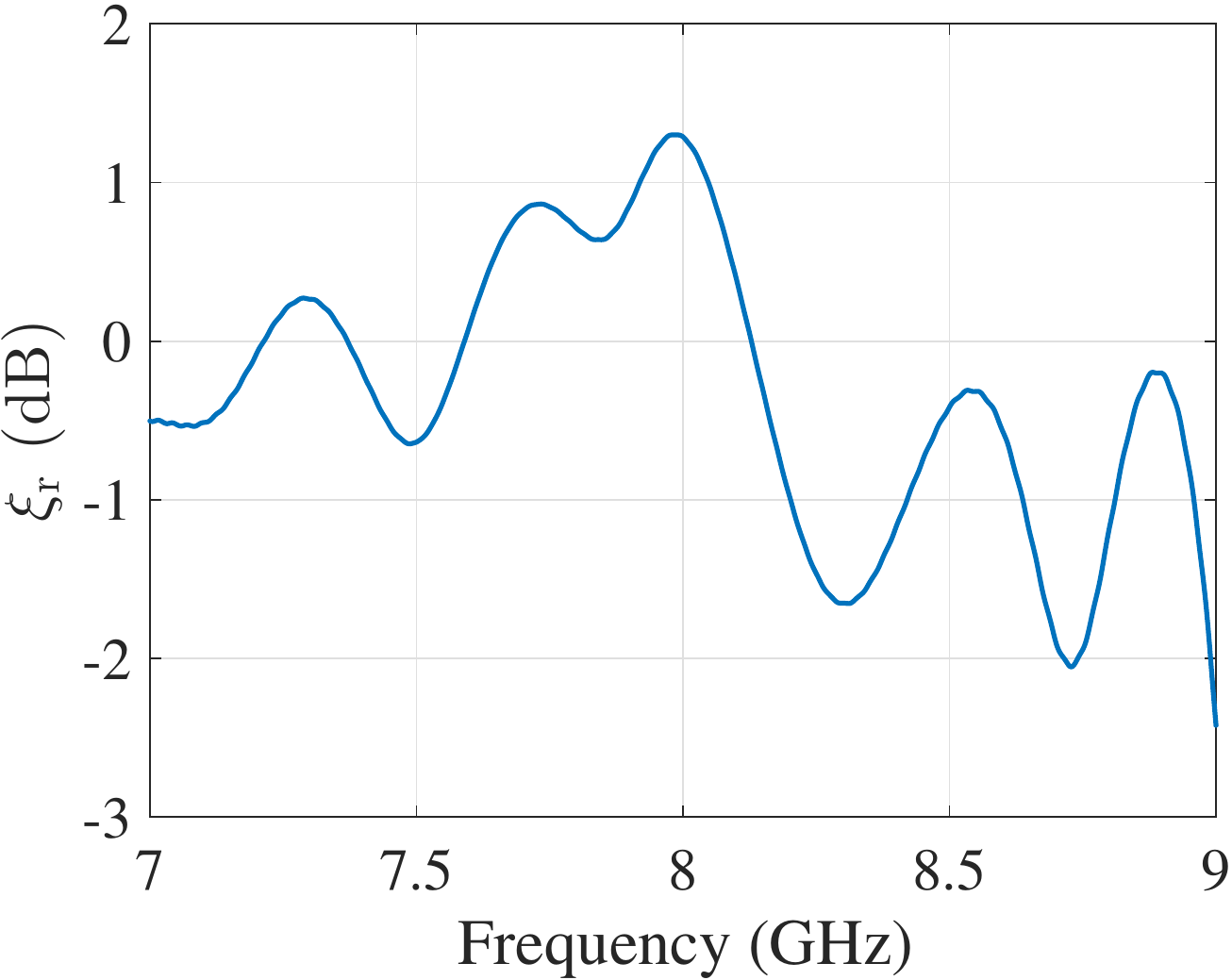, width=0.46\columnwidth} 
		\label{fig7b} }  \\
			\subfigure[]{
		\epsfig{file=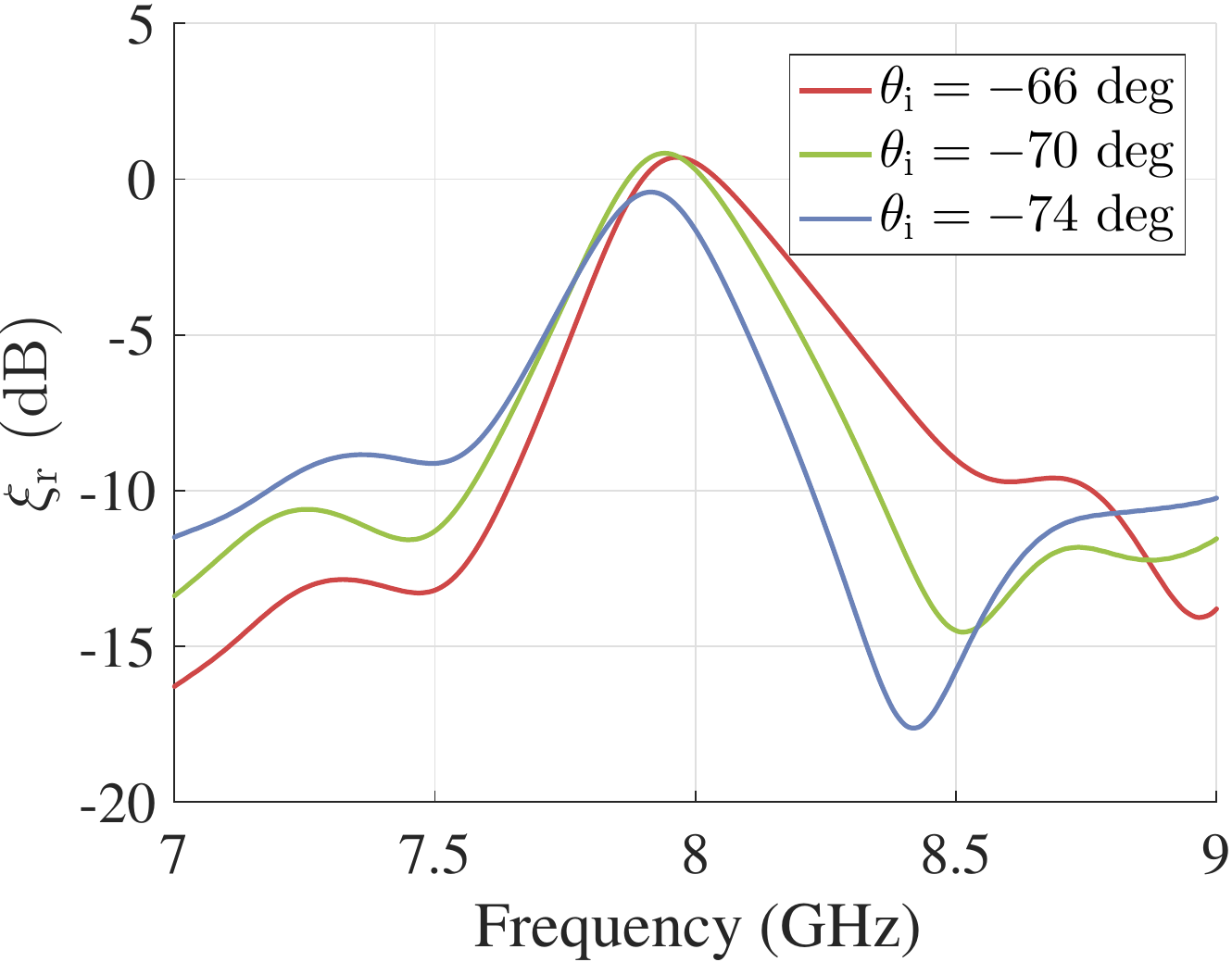, width=0.46\columnwidth}  
		\label{fig7c} }
			\subfigure[]{
		\epsfig{file=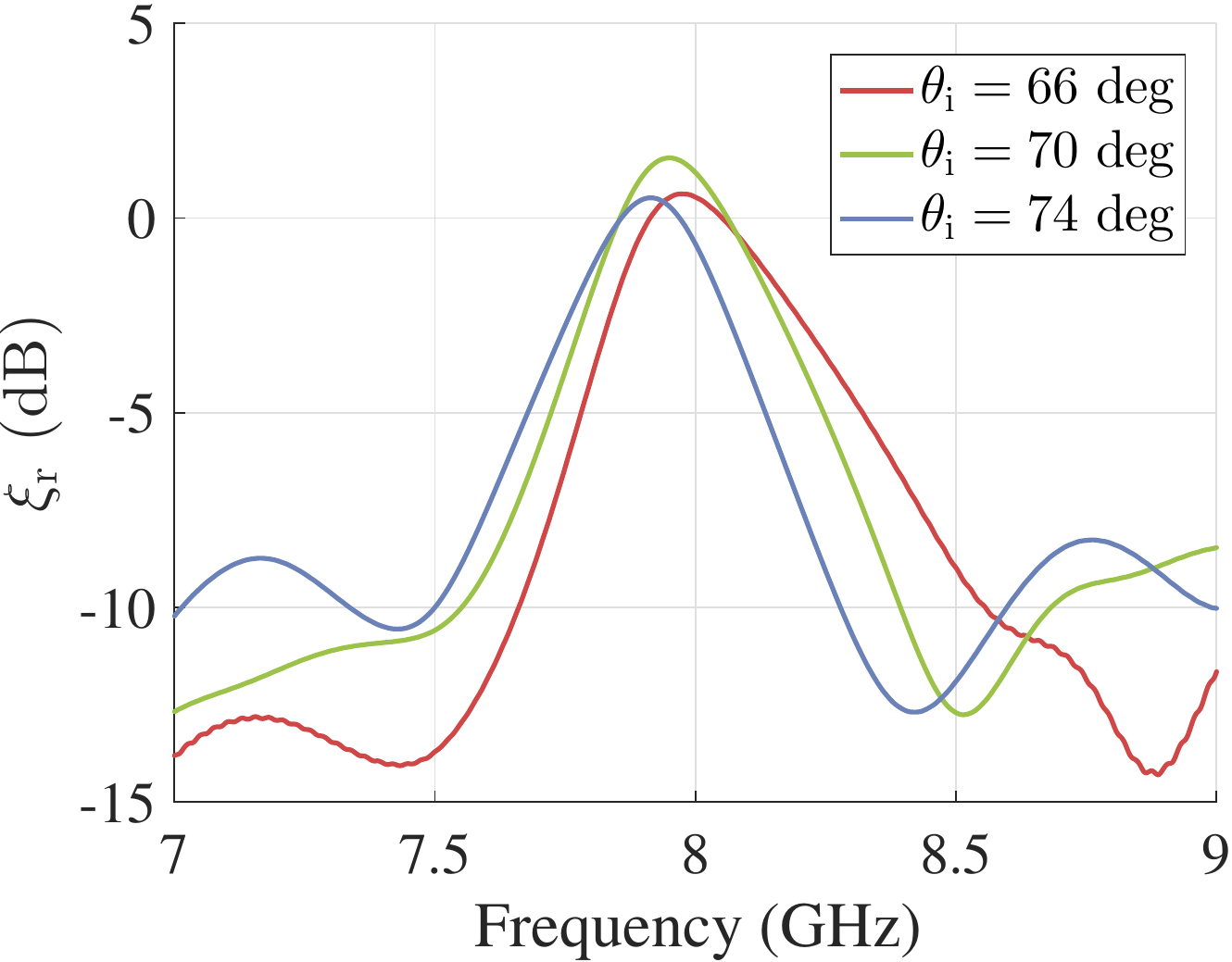, width=0.46\columnwidth}  
		\label{fig7d} }
         \caption{(a) Illustration of the experimental set-up (top view). Measured reflection efficiency of the  metasurface when illuminated (b) normally, (c)  at angles near $\theta_{\rm i}=-70^\circ$, and (d) at angles near $\theta_{\rm i}=+70^\circ$.
	}\label{fig7}
\end{figure}
The sample was illuminated by a quad-ridged horn antenna (11~dBi gain at 8~GHz) which was located at a distance of 5.5~m (about $147\lambda$) from the sample and also played the role of a receiving antenna. To filter out parasitic reflections from the chamber walls and  antenna cables, the conventional time gating post-processing technique was exploited. In order to find the reflection efficiency $\xi_{\rm r}$ of the sample, defined as the ratio of the incident and reflected power densities, we measured the received signal from the sample and then normalized it by the corresponding received signal from a reference uniform aluminium plate of the same cross section size. 

The measured reflection efficiency of the sample when it is illuminated normally (from port~2; $\theta_{\rm i}=0^\circ$) is shown in Fig.~\ref{fig7b}. As it was discussed above, under the normal illumination the sub-wavelength periodicity of the metasurface allows only the $n=0$ propagating channel (specular reflection). Therefore, at the resonance frequency 8~GHz and below, the structure reflects back nearly all the incident power. Interestingly, at 8~GHz the reflection efficiency reaches 1.288~dB (135\% of the power). This non-physical result is a consequence of the normalization error which wrongly implies that the currents induced at the metasurface and the reference metal plate should be equally uniform. More details on the proper normalization procedure in this case can be found in \cite{suppl}. 
Figures~\ref{fig7c} and \ref{fig7d} depict the reflection efficiency when the sample was illuminated at angles near the directions of the other two open channels $\pm 70^\circ$.  As is seen, there are strong peaks of reflection at 8~GHz: $\xi_{\rm r}=0.3$~dB and $\xi_{\rm r}=1.17$~dB for incidence at $-70^\circ$ and $+70^\circ$, respectively. These results, with admissible measurement error level of 1~dB  for such type of measurements, well confirm the high reflection efficiency of all channels predicted by the full-wave simulations.   

\section{Three-channel power splitter} \label{splitter}
To further illustrate the versatility of the proposed multi-channel
paradigm, we synthesize a three-channel power splitter. The response of the splitter can be considered as a combination of  two basic functionalities (anomalous reflections at two angles) dictated by (\ref{m2}) and  (\ref{m21}). Beam splitters have been studied intensively in the form of diffraction gratings (see e.g., \cite{splitter1,splitter2}), however, to the best of our knowledge, the possibility to tailor the proportion of power divided into different directions was not reported. Usually, the power is split equally between all channels. In contrast, our synthesis technique can be used to  design a high-efficiency beam splitter with arbitrary   energy distribution between the ports and arbitrary transmission phases. In the following example, we target a metasurface that under normal incidence splits incident energy towards $-70^\circ$, $0^\circ$, and $+70^\circ$ angles with the proportion 50:0:50. A metasurface with   the same functionality and designed using the same design approach was recently proposed  in~\cite{splitter3} (only as a theoretical impedance sheet), but no physical implementation was demonstrated.  

In our design, normally incident plane waves (from port~2) are split between 
ports~1 and~3 [see Fig.~\ref{fig1b}] without any reflection into port~2, i.e., port~2 is matched to the free-space wave impedance.  Considering TE-polarized modes, we write the total tangential electric $E_t$ and magnetic $H_t$ fields on the 
reflecting surface at $z=0$ as
\begin{equation}
\begin{array}{c}\displaystyle
E_t = E_{\rm i}+E_{\rm r1}e^{-jk_{\rm r} \sin \theta_{\rm r} x+j\phi_1}+
E_{\rm r3}e^{jk_{\rm r} \sin \theta_{\rm r} x+j\phi_3},\\ \displaystyle
H_t = \frac{E_{\rm i}}{\eta}-\frac{\cos\theta_{\rm r}}{\eta}
\left(E_{\rm r1}e^{-jk_{\rm r} \sin \theta_{\rm r} x+j\phi_1}+E_{\rm r3}e^{jk_{\rm r} \sin \theta_{\rm r} x+j\phi_3}\right), \displaystyle
\end{array}
\label{tang}
\end{equation}
where $E_{\rm r1}$, $E_{\rm r3}$ and $\phi_1$, $\phi_3$ refer to
the amplitudes and phases of the reflected waves at ports~1 and 3,
respectively, and $k_{\rm r}=k_{\rm i}$ is the wavenumber of the reflected waves.
Equal power division occurs when
$E_{\rm r1}=E_{\rm r3}=E_{\rm i}/\sqrt{2\cos\theta_{\rm r}}$. Interestingly, if phases $\phi_1=\phi_3=0\degree$   are chosen, the surface impedance $Z_{\rm s}=E_t/H_t$ becomes purely real, meaning that the  the Poynting vector has only the normal component (negative or positive in different locations) at the impedance boundary. Designing a metasurface to implement this scenario is not straightforward with our approach which implies optimizing  unit-cell elements to follow the imaginary part of the required surface impedance.    Therefore, in our example, we choose  reflection phases  $\phi_1=180\degree$ and $\phi_3=0\degree$ to simplify the metasurface design.
In this case, the corresponding surface impedance  is expressed as
\begin{equation}
Z_{\rm s}=\frac{\eta}{\sqrt{\cos\theta_{\rm r}}}
\frac{\sqrt{\cos\theta_{\rm r}}+j\sqrt{2}\sin(k_{\rm r} \sin \theta_{\rm r} x)}
{1-j\sqrt{2\cos\theta_{\rm r}}\sin(k_{\rm r} \sin \theta_{\rm r} x)},
\label{zs_splitter}
\end{equation}
which is a periodic function with the period
$D=\lambda/\sin\theta_{\rm r}$. Notice that this periodicity is the double of that in the previous example of an isolating mirror. 
The surface impedance is plotted in Fig.~\ref{fig8a}.
\begin{figure}[t]
  \subfigure[]{
	 \epsfig{file=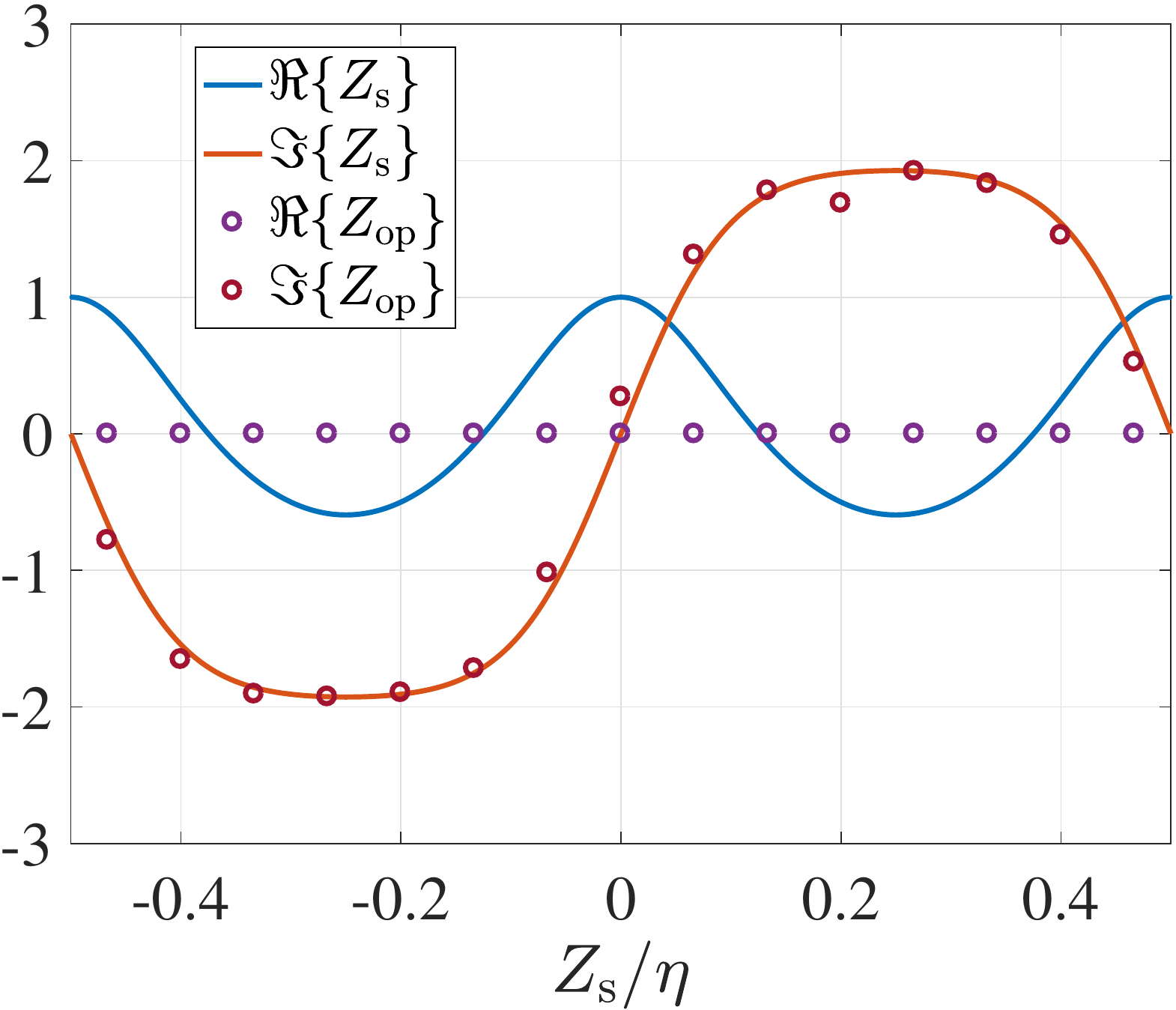,width=0.46\columnwidth}
	 \label{fig8a}}
	\subfigure[]{
	 \epsfig{file=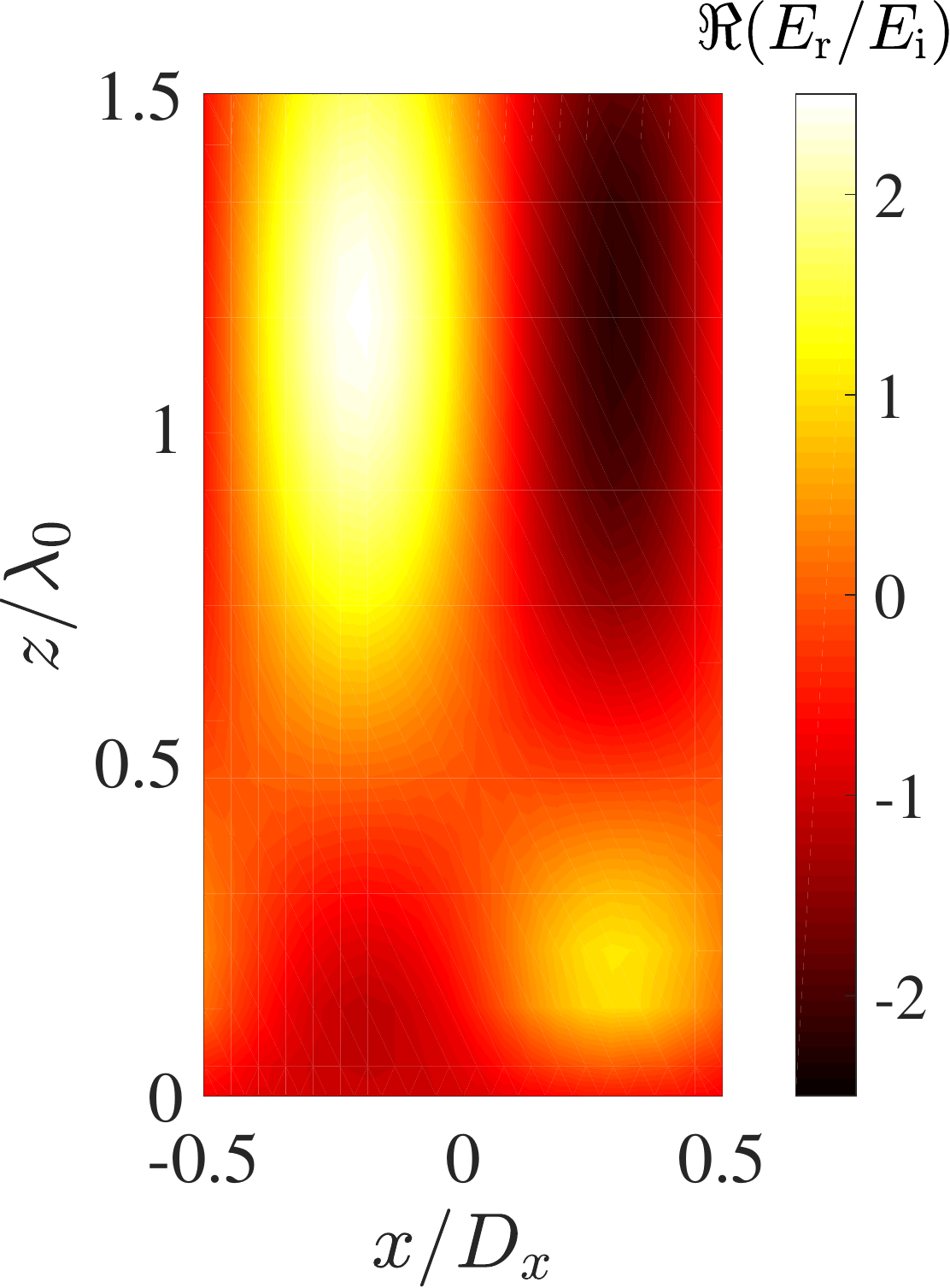,width=0.4\columnwidth}
	 \label{fig8b}}
 \caption{
A 50:0:50 power splitter. (a) Surface impedance dictated by (\ref{zs_splitter}) shown with solid lines and optimized surface impedance shown with circle marks.  (b) A snapshot of the scattered normalized  electric field (real value) in the proximity of the splitter located at $z=0$. The data corresponds to the splitter implementation shown in Fig.~\ref{fig9a}. }
 \label{fig8}
\end{figure}
The surface resistance $\Re\{Z_{\rm s}\}$ oscillates between
positive and negative values, associated with locally lossy 
and active properties, respectively. Averaged over the period,
the surface is in  overall lossless.

Instead of an inhomogeneous surface with locally active and lossy regions,
the reflecting surface may be implemented as a strongly non-local
lossless metasurface based on  energy channelling along the
surface~\cite{ana,last}. Indeed, the surface impedance~(\ref{zs_splitter}) has been derived assuming that there are only three plane waves propagating above the metasurface: A single incident and two reflected ones. This assumption is very limiting and in general can be relaxed by allowing evanescent waves generation at the interface. While they would not contribute to the far-field radiation, decaying quickly near the interface, their contribution to the total fields (\ref{tang}) at $z=0$ would ensure purely imaginary surface impedance~$Z_{\rm s}$ at each point of the metasurface. Although analytical determination of the required set of evanescent waves is a complicated task, the solution can be obtained exploiting the concept of leaky-wave antennas~\cite{ana}. 
In this implementation, the energy received over the lossy
region (where $\Re\{Z_{\rm s}\}>0$) is carried by a surface
wave propagating along the $x$-axis before being radiated into space in the active
region of the surface (where $\Re\{Z_{\rm s}\}<0$). In what follows, we discard the real part of  impedance~(\ref{zs_splitter}) and numerically optimize~\cite{HFSS} the imaginary part of the impedance to ensure the desired amplitudes of the reflected \textit{plane} waves.
The optimized impedance profile (discretized into 15  elements with a uniform impedance) is shown by  circle markers  in Fig.~\ref{fig8a}.
The numerically optimized impedance profile (the imaginary part) turns out to closely follow that  of the active-lossy impedance dictated by (\ref{zs_splitter}). 

Fig.~\ref{fig8b} depicts numerical results for the real part (the instantaneous value) of the  normalized electric field  reflected by the metasurface with the optimized impedance profile.  
An interference pattern between two plane waves propagating
in the $xz$-plane towards the $-70\degree$ and $+70\degree$ directions 
with a $180\degree$ phase difference is observed. As is expected, at the interface a set of evanescent fields is generated. The incident wave is split equally between ports~1 and 3 with 49.8\% of incident power channelled into each direction.

Following the same implementation approach as for the three-channel isolating mirror, we use rectangular patches over a metallic ground plane. Ten equidistant patches with the same width of  $2.69$~mm and different lengths were placed in each unit cell. As an initial estimation, the length of each patch was chosen in order to ensure the reflection phase described by the imaginary part of the optimized impedance when the patch is placed in a homogeneous array. After that, we built the unit cell consisting of ten patches and accurately tuned the lengths by using post-optimization for obtaining the best performance. The final lengths of the patches listed along the $+x$-direction are $4.9$, $7.4$, $9.2$, $10.6$, $11$, $18$, $11.6$, $11.5$, $10.6$, and $11.2~{\rm mm}$ [see Fig.~\ref{fig9a}]. 
\begin{figure}[h]
	\centering
	\subfigure[]{
		\epsfig{file=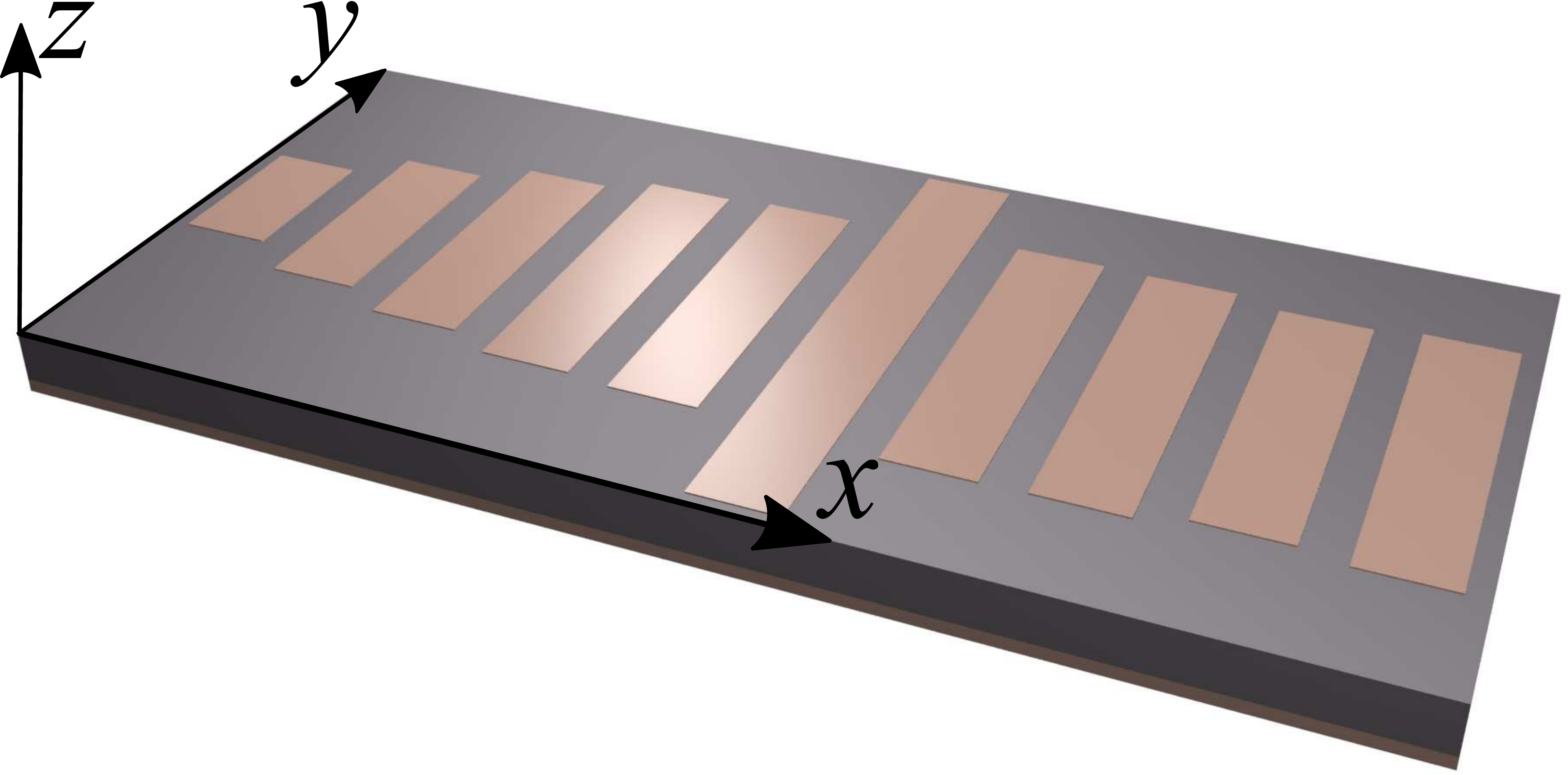, width=0.45\columnwidth}  
		\label{fig9a} }
	\subfigure[]{
		\epsfig{file=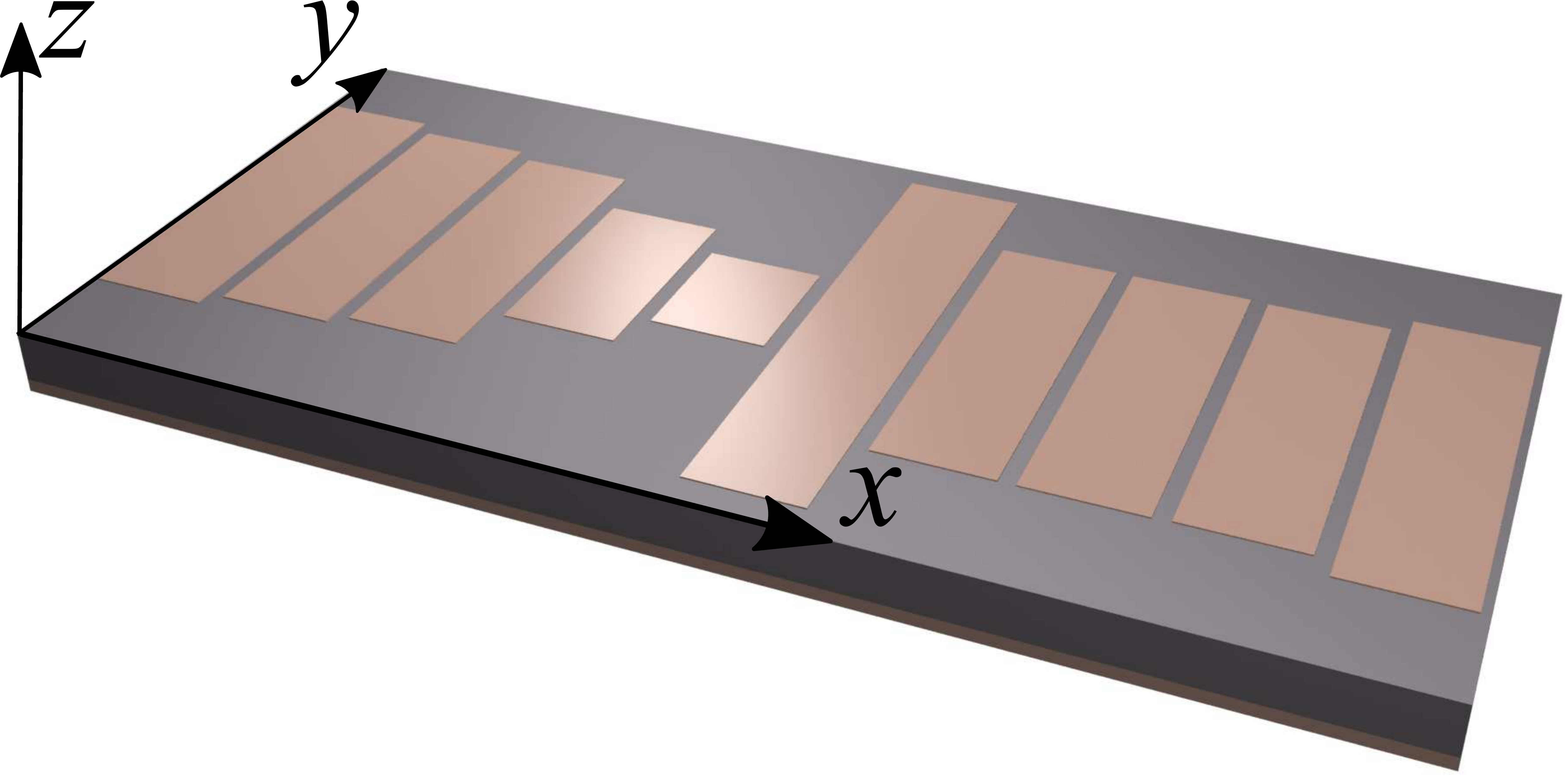, width=0.45\columnwidth} 
		\label{fig9b} }  
	\caption{Schematic representation of single unit cell of designed (a) three-channel power splitter and (b) five-channel retro-reflector.
 }\label{fig9}
\end{figure}
The dimensions of the unit cell  are $D=\lambda/\sin 70^\circ=39.91$~mm and $\lambda/2=18.75$~mm along the $x$-axis and $y$-axis, respectively. The substrate material and patches material are the same as in the previous example. The numerical results indicate that normally incident power is split by the metasurface towards $-70^\circ$ (port~3), $0^\circ$ (port~2), and $+70^\circ$ (port~1) with the proportion  \mbox{$49.71\%:1.06\%:45.74\%$}. About 3.49\% of incident power is absorbed in the metasurface due to material losses. 
When the splitter is illuminated from port~1 or 3, it is not matched to free space, and energy is split in all three ports with the proportion of \mbox{$23.26\%:45.74\%:28.51\%$} (when illuminated from port~1)  or \mbox{$25.19\%:49.71\%:23.26\%$} (when illuminated from port~3). 
This result is not surprising and in agreement with  the basic multi-port network theory which states that lossless passive three-port splitters matched at all  ports cannot exist (since it would contradict  the condition of  unitary scattering matrix $S_{k i}^* \,S_{k j} = \delta_{i j}$~\cite{pozar}). 

Next, we experimentally verify operation of the metasurface  in an anechoic chamber. The metasurface sample consists of $11 \times 14$ unit cells along the $x$ and $y$ axes, respectively. The overall dimensions of the sample along the $x$ and $y$ axes are $11.7\lambda=439$~mm and $7\lambda=262.5$~mm. 

First, the sample was illuminated from the normal direction ($\theta_{\rm i}=0^\circ$) by a horn antenna as is shown in Fig.~\ref{fig7a}, and the  reflected signal was measured by the same antenna. Similarly to the experiment in the previous section, the signal was normalized by that of a reference uniform aluminium plate of the same size. The measured level of retro-reflection~$\xi_{\rm r}$ along direction~$\theta_{\rm i}=0^\circ$ is depicted in Fig.~\ref{fig10a}. 
\begin{figure}[h]
	\centering
	\subfigure[]{
		\epsfig{file=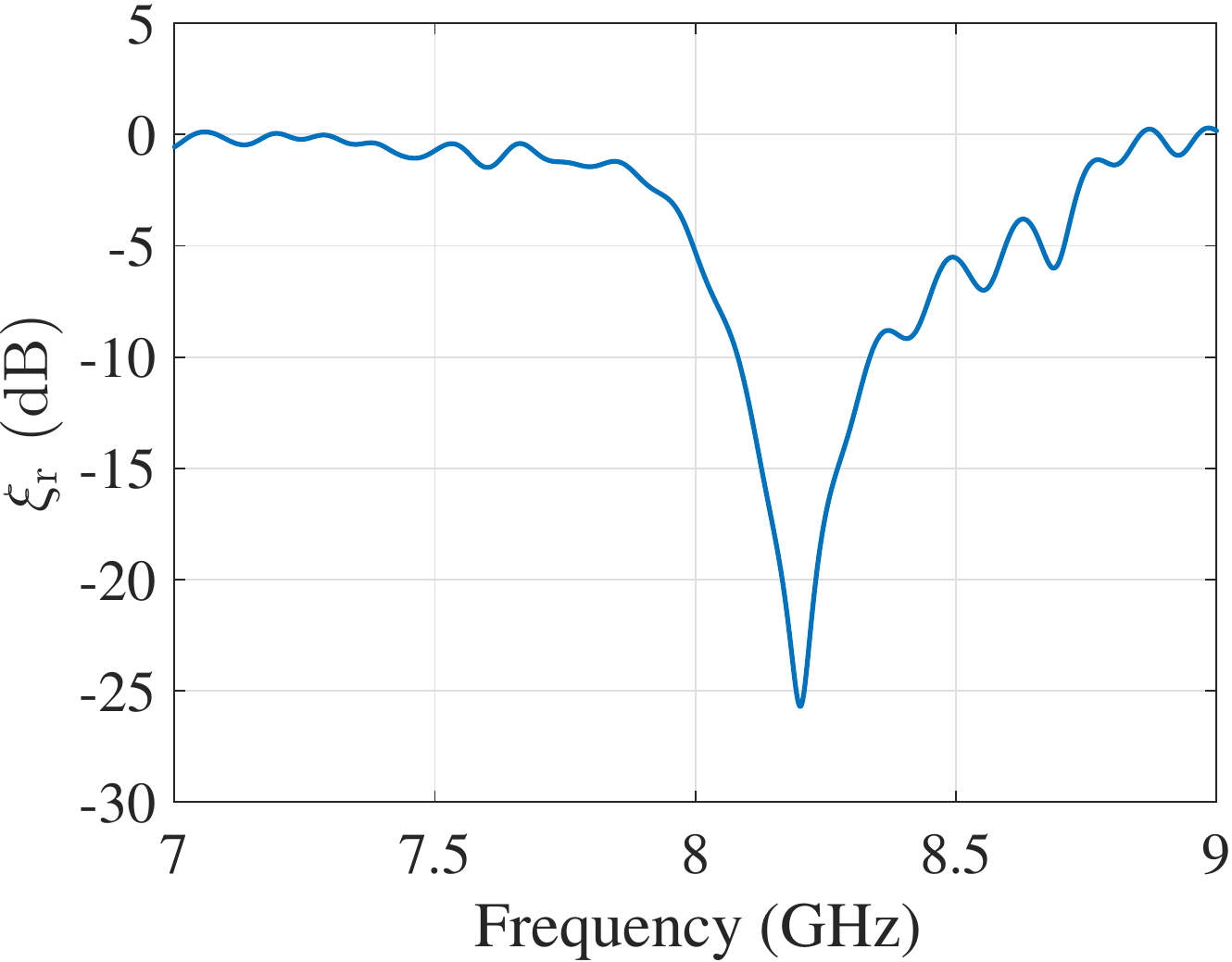, width=0.46\columnwidth} 
		\label{fig10a} } 
			\subfigure[]{
		\epsfig{file=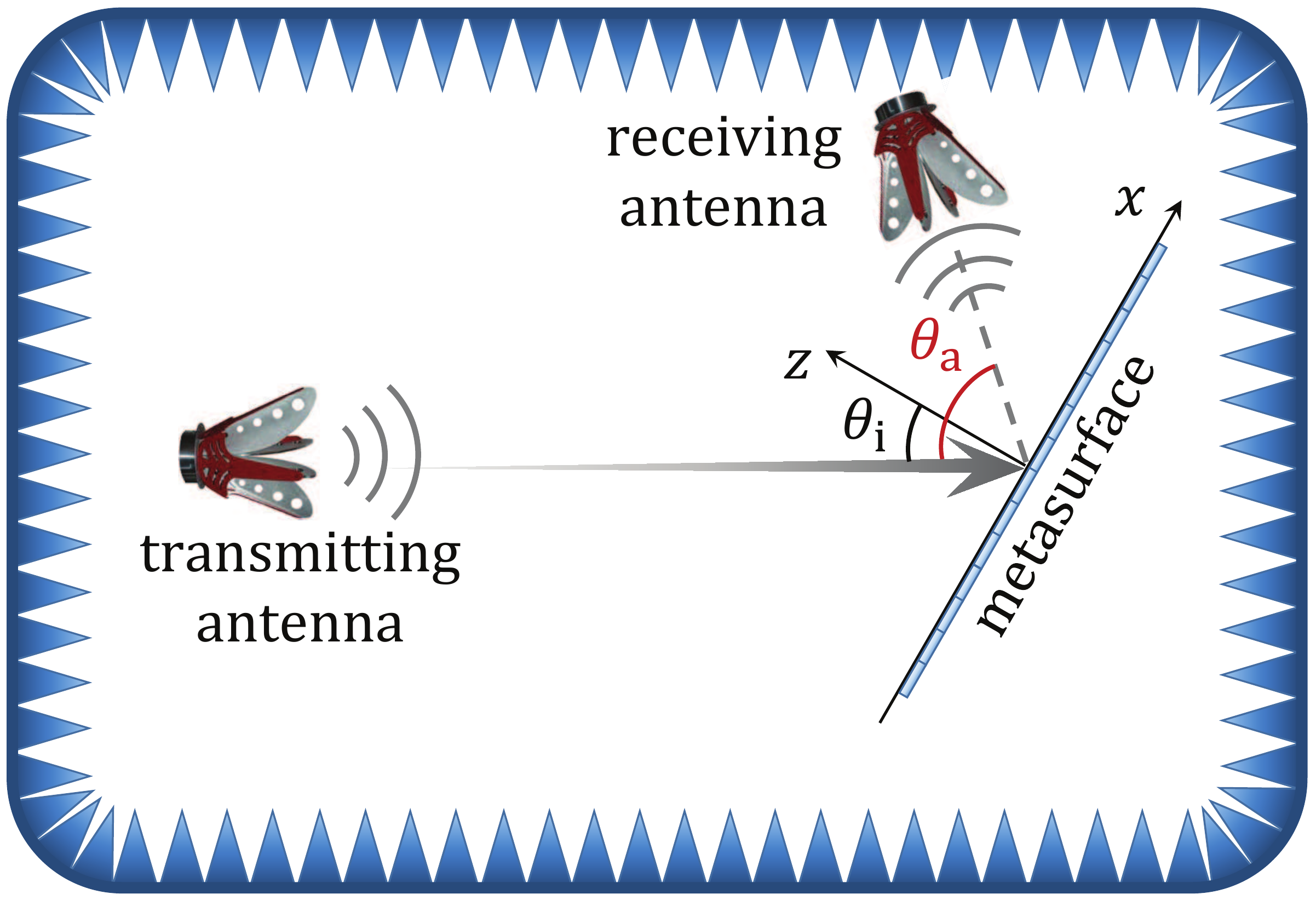, width=0.46\columnwidth}  
		\label{fig10b} }   \\
			\subfigure[]{
		\epsfig{file=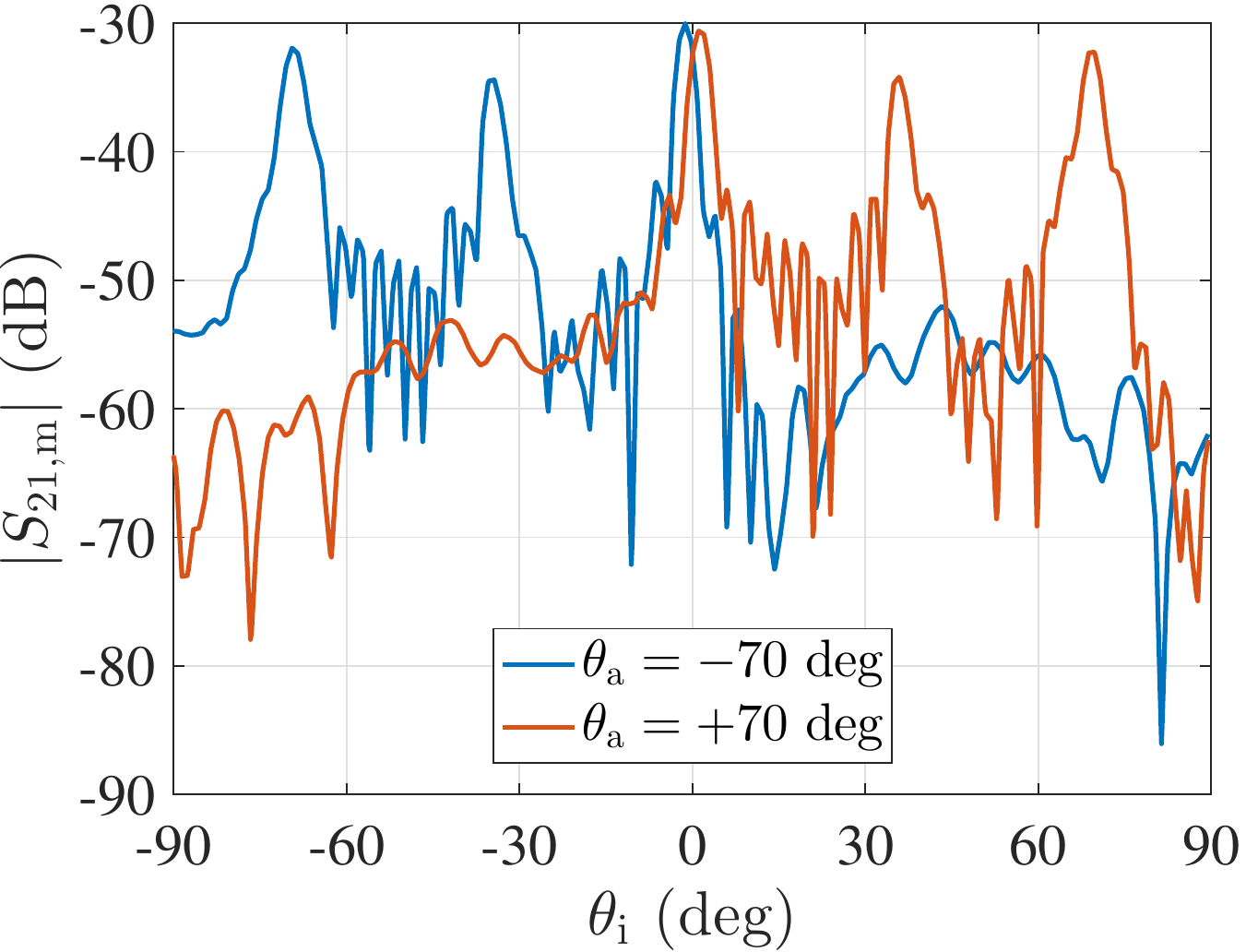, width=0.46\columnwidth}  
		\label{fig10c} }
			\subfigure[]{
		\epsfig{file=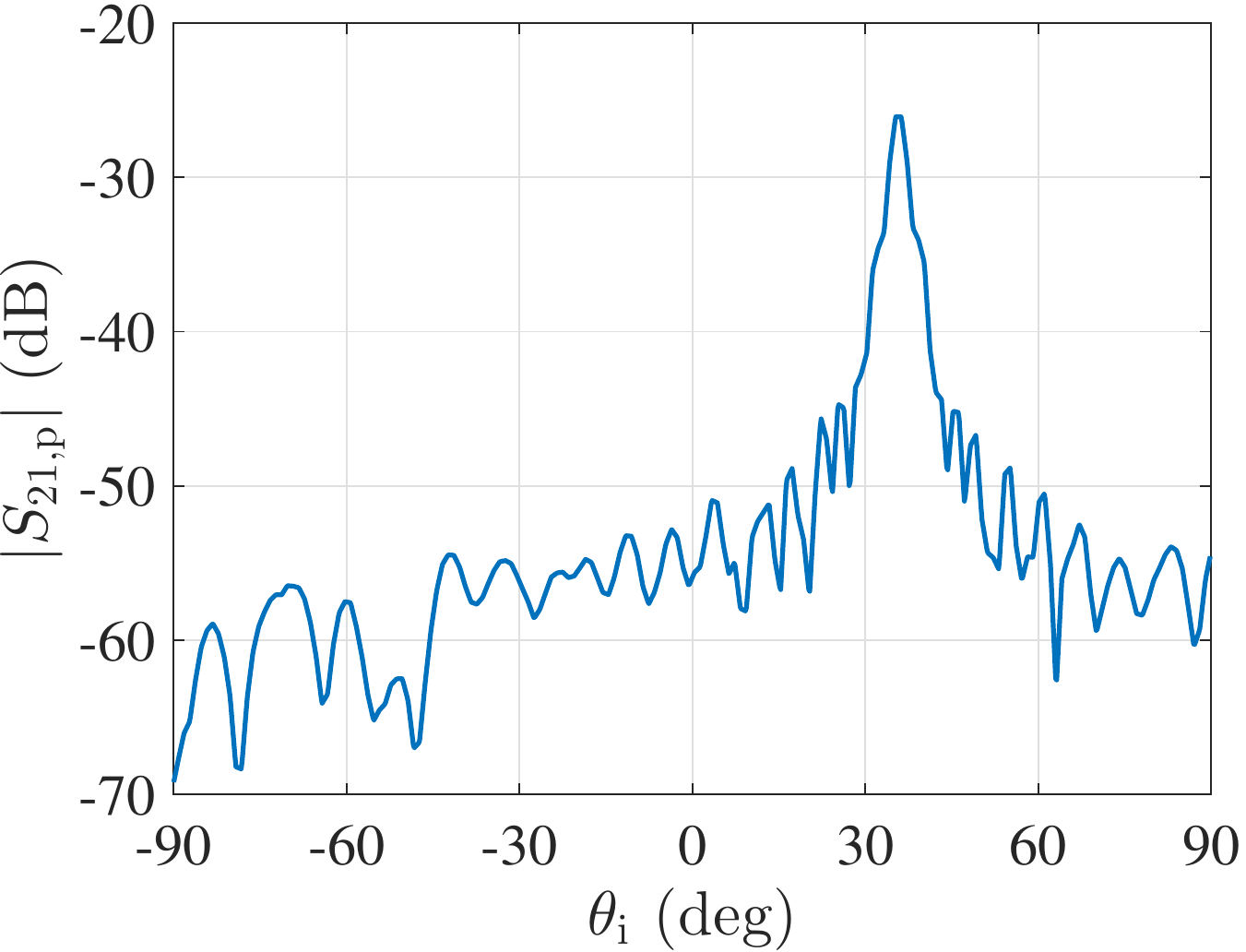, width=0.46\columnwidth}  
		\label{fig10d} }
         \caption{(a) Efficiency of retro-reflection from the power splitter under normal illumination. The splitter is matched for normal incidence at 8.2~GHz. (b) Illustration of the experimental set-up for the power splitter (top view).  (c) Signals measured by the receiving antenna versus the orientation angle of the metasurface $\theta_{\rm i}$ for two  positions of the receiving antenna. (d)  Signal  from a reference metal plate ($\theta_{\rm a}=+70^\circ$). 
	}\label{fig10}
\end{figure}
At the resonance frequency of 8.2~GHz, the measured curve has a deep drop, corresponding to reflected power of 0.3\%. This result is in excellent agreement with the simulated data and confirms that the power splitter is well matched from port~2. A slight shift of the resonance frequency in the experiment can  be explained by manufacturing errors and the tolerance of the substrate permittivity value. 

Next, it is important to verify that the splitter in fact reflects normally incident power equally towards $+70^\circ$ and $-70^\circ$ directions. For this purpose, we add a se\-cond (receiving) antenna to the experimental setup [see Fig.~\ref{fig10b}].  The position of the se\-cond antenna is determined by  the angle $\theta_{\rm a}$. The distances from the metasurface center to the receiving and transmitting antennas were~2.1~m and 5.4~m, respectively. Figure~\ref{fig10c} depicts the  signal measured by the receiving antenna $|S_{21, {\rm m}}|$ for two different positions of the antenna (when $\theta_{\rm a}=+70^\circ$ and $\theta_{\rm a}=-70^\circ$) versus the  incidence angle $\theta_{\rm i}$. The red curve ($\theta_{\rm a}=+70^\circ$) has three distinct peaks. The central peak at $\theta_{\rm i}=0^\circ$ corresponds to the case of normal incidence (from port~2) and strong reflection towards port~1 ($\theta_{\rm r}=+70^\circ$). The  peak at   $\theta_{\rm i}=+70^\circ$ corresponds to excitation of the metasurface from port~3 and reflection towards port~2 ($\theta_{\rm r}=0^\circ$). The  peak at   $\theta_{\rm i}=+35^\circ$ appears due to specular reflection from the metasurface when it is illuminated not from its main ports. Three peaks of  the blue  curve for the case   of $\theta_{\rm a}=-70^\circ$ can be explained likewise. Interestingly, the central peaks  on the red and blue curves do not occur exactly at $\theta_{\rm i}=0^\circ$ and there is a small shift between them. This result is expected and means that when the power splitter is illuminated at a small oblique angle, the reflection towards one direction increases at the expense of the decrease of reflection in the other direction. The total efficiency (sum of the normalized reflected power towards $+ 70^\circ$ and $-70^\circ$ directions) is maximum exactly when  $\theta_{\rm i}=0^\circ$.

In order to calculate the power reflected by the metasurface towards $\theta_{\rm r}=+70^\circ$ and $\theta_{\rm r}=-70^\circ$, we perform an additional  measurement with a reference aluminium plate (position of the receiving antenna was at $\theta_{\rm a}=+70^\circ$). As is seen from Fig.~\ref{fig10d}, a specular reflection peak is detected by the antenna when the metal plate is oriented at $\theta_{\rm i}=+35^\circ$. Following the procedure reported in~\cite{ana}, we calculate the reflection efficiency $\xi_{\rm r}$ of the power splitter towards $\theta_{\rm r}=+70^\circ$ (for the red curve) and $\theta_{\rm r}=-70^\circ$ (for the blue curve) as
\begin{equation}
\xi_{\rm r}=\frac{1}{\xi_{0}} \frac{|S_{21, {\rm m}}(\theta_{\rm i}=0^\circ)|}{|S_{21, {\rm p}}(\theta_{\rm i}=+35^\circ)|}.
\label{eff}
\end{equation}
Here $\xi_{0}$ is a correction factor  which gives the ratio between
the theoretically calculated signal amplitudes from an ideal power splitter (of the same size and made of lossless materials) and a perfect conductor plate~\cite{ana}. Using this formula, the calculated reflection efficiencies towards $\theta_{\rm r}=+70^\circ$   and $\theta_{\rm r}=-70^\circ$ directions were found to be nearly equal  to $\xi_{\rm r}^{+70^\circ} \approx \xi_{\rm r}^{-70^\circ}\approx -3.17$~dB.  In the linear scale and expressed in terms of power, the reflection efficiency towards each direction (ports~1 and 3) is about 48\%. This result is in good agreement with the simulated data (45.74\% and 49.71\%).

\section{Five-channel retro-reflector} \label{5port}
In this section, we demonstrate that the concept of multi-channel flat reflectors can be extended to devices with more than three ports. In what follows, we synthesize a mirror which fully reflects incident waves back from five different directions. 
Since such functionality requires fast variations of the surface impedance over the sub-wavelength scale (due to excitation of required additional evanescent fields for satisfying the boundary condition), it is unattainable with conventional blazed gratings whose grooves are of the wavelength size. 
Designing  conventional  gratings with sub-wavelength-sized grooves is   impractical with today's nano-fabrication technologies.

In the  case of the three-channel retro-reflector described above,  engineering back reflections in one port automatically ensured proper response of the reflector in the other two ports (due to reciprocity and negligible dissipation). However, when the number of ports increases to five, one should design a metasurface with prescribed response for several different illuminations.

 The directions of the five ports cannot be chosen   arbitrary and  can be determined using Floquet-Bloch analysis: $\theta_{{\rm r} n}=\arcsin (\sin \theta_{\rm i}+ n \lambda/D)$, where $n=0, \pm 1, \pm 2$ is the number of the harmonic and $\theta_{{\rm r} n}$ is the reflection angle corresponding to the $n$-th harmonic.
There are two scenarios of  retro-reflection in five ports. In the first scenario, the periodicity of the reflector $D$ is chosen to satisfy inequality $2\lambda<D<3\lambda$. In this case, all the five ports are ``open'' when the reflector is illuminated from any of them. The directions of the five  ports are given by $\theta_{{\rm r} n}=\arcsin ( n \lambda/D)$, where  $n=0, \pm 1, \pm 2$.

In the second scenario, the reflector periodicity  $D$ is chosen so that   $\lambda<D<2\lambda$. Here, at normal illumination $\theta_{\rm i}=0^\circ$, only three ports are open: At angles $\theta_{{\rm r} 0}=0^\circ$ and $\theta_{{\rm r} \pm 1}=\pm \arcsin ( \lambda/D)  $ [see Fig.~\ref{fig2b}], where subscript indices correspond to the harmonic number $n$. The same ports remain open also in the case when $\theta_{\rm i}=\pm \arcsin ( \lambda/D)$ [see  Fig.~\ref{fig11a} where  $D \approx 1.064\lambda$ is chosen as an example]. However, illuminated from  $\theta_{\rm i}=\pm \arcsin (0.5 \lambda/D)$, the metasurface has only two open ports corresponding to  $\theta_{{\rm r} -1,0}=\pm \arcsin (0.5 \lambda/D)$, as is shown in Fig.~\ref{fig11b}. Thus, although the retro-reflector has five channels, three of its channels are isolated from two others.

Both scenarios of  the retro-reflector provide similar response for the illuminations of the five channels. For our design  we use the second scenario since it requires to ensure retro-reflection only in three ports (response in others will be automatically satisfied due to reciprocity). We choose   periodicity $D\approx 1.064\lambda$ so that the five ports are directed at $0^\circ$, $\pm28^\circ$, and $\pm70^\circ$ from the normal.

\begin{figure}
	\centering
	\subfigure[]{
		\epsfig{file=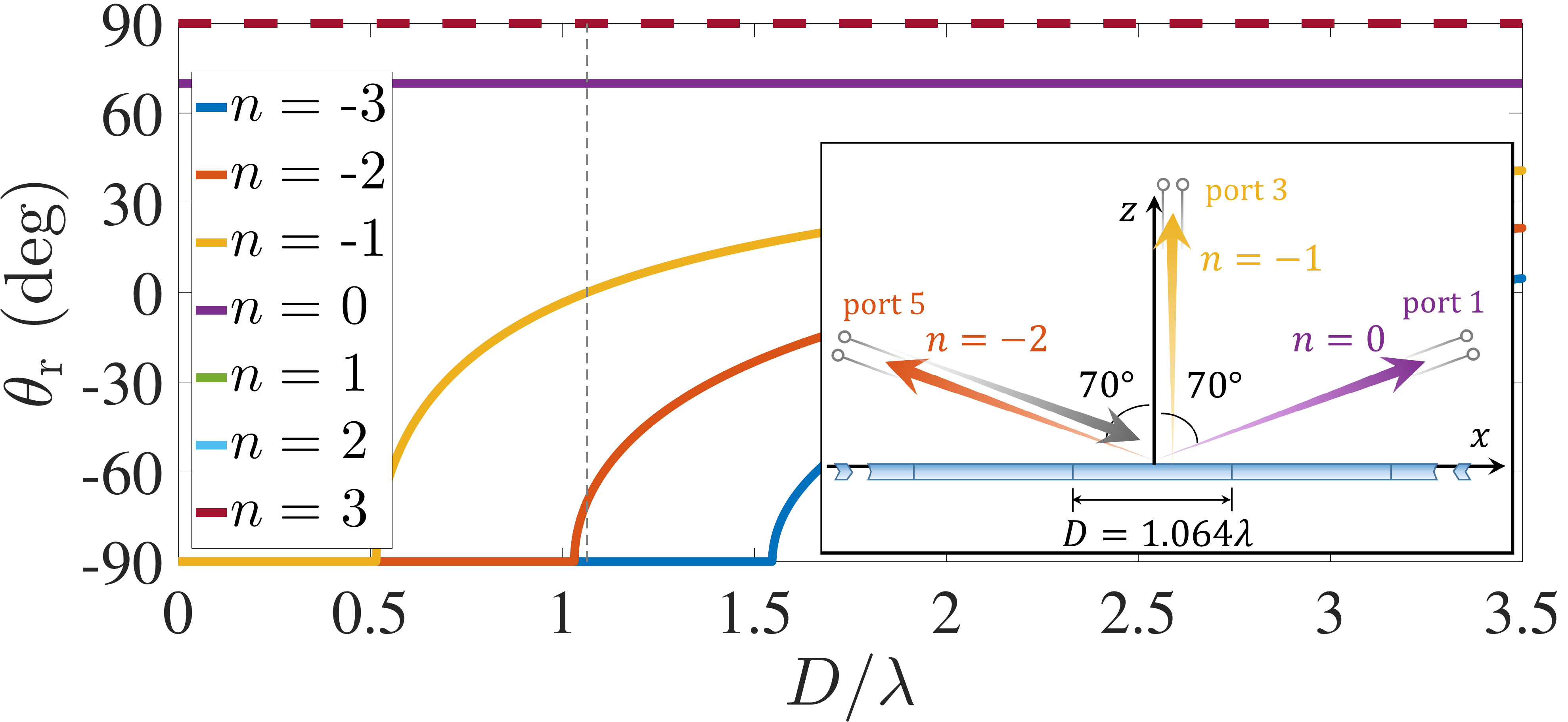, width=0.96\columnwidth}  
		\label{fig11a} }\\
	\subfigure[]{
		\epsfig{file=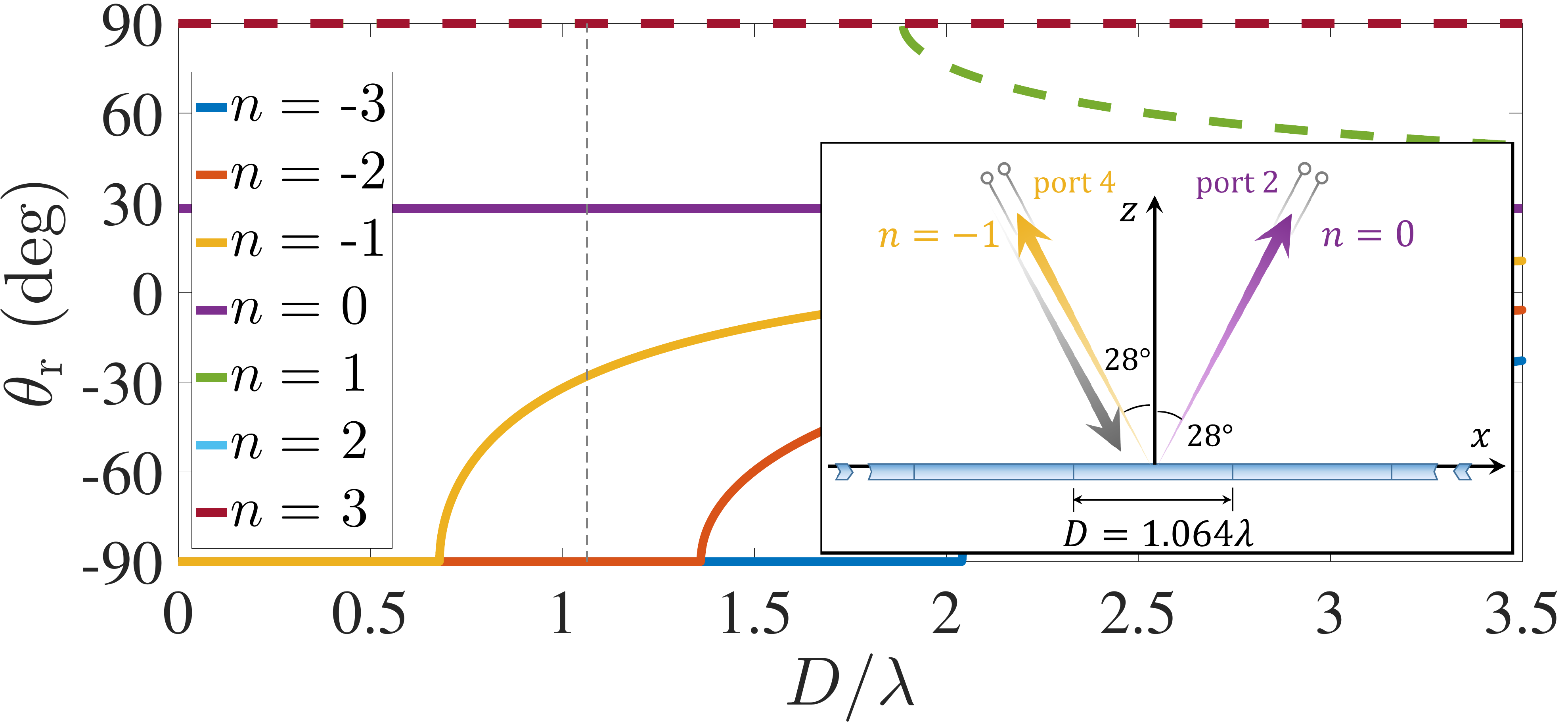, width=0.96\columnwidth} 
		\label{fig11b} }  
         \caption{The reflection angles of different Floquet harmonics for a five-channel retro-reflector. Here, the reflector periodicity $D \approx 1.064\lambda$ is chosen.   Illumination is at (a)   $\theta_{\rm i}=70^\circ$ and (b) $\theta_{\rm i}=28^\circ$. The grey dashed line indicates the periodicity of the reflector.  For clarity, only seven harmonics are shown.
	}\label{fig11}
\end{figure}

It can be shown that  an ideal five-channel retro-reflector illuminated from either channel necessarily generates evanescent waves.  Analytical determination of these required evanescent waves is a complex problem and could  be a subject for a separate study. Next, we firstly design a metasurface for   retro-reflection  when illuminated  from $\theta_{\rm i}=\pm 28^\circ$ directions and then exploit numerical  post-optimization to gain proper response in the other channels. 

As the first approximation, we synthesize the metasurface to fully reflect incident waves from $\theta_{\rm i}=+28^\circ$ at angle $\theta_{\rm r}=-28^\circ$. This case is similar to that considered above in Section~\ref{three} and, therefore, the impedance profile can be calculated  using Eq.~(\ref{eq:4}). The obtained metasurface rigorously works as a retro-reflector from ports~2 and 4 [see ports notations in Fig.~\ref{fig11}], while it produces  parasitic energy coupling  to  ports~1 and 5 when excited from the normal direction (port~3). Next, we numerically optimize~\cite{HFSS} the obtained impedance profile to ensure maximized back reflection (retro-reflection) in all the ports. The optimized unit cell of the metasurface is shown in Fig.~\ref{fig9b} and contains ten patches with the following lengths listed along the $+x$-direction: $12.3$, $11.2$, $10.4$, $6.8$, $4.1$, $16.9$, $11.1$, $11.1$, $11.0$, and $12.3~{\rm mm}$. The unit cell dimensions are the same as those of the three-channel power splitter considered above.  The width of the patches is 3.49~mm. The numerically obtained results can be represented by the following matrix (at 8~GHz)
\begin{equation}
(|S|^2)_{i j} = \begin{pmatrix} 
0.796  & 0     & 0.136  & 0     & 0     \\ 
0      & 0.815 & 0      & 0.156 & 0     \\
0.136  & 0     &  0.784 & 0     & 0.064 \\
0      & 0.156 &  0     & 0.807 & 0     \\
0      & 0     &  0.064 & 0     & 0.873 
     \end{pmatrix},
     \label{sm} 
\end{equation}
where $(|S|^2)_{i j}$ characterizes the ratio of power reflected into the $i$-th port when the metasurface is illuminated from the $j$-th port [see  definitions of the ports in Fig.~\ref{fig11}]. As is seen, for illumination from either port, the metasurface effectively reflects energy back in the incident direction with the average efficiency of $80\%$, i.e. $(|S|^2)_{i i} \approx 0.8$ for $i=1, 2\dots5$. The non-ideal operation can be improved further by increasing discretization of the unit cell.

Next, we fabricated a metasurface sample with the same dimensions as those presented in Section~\ref{splitter}. The experimental setup was identical to that shown in Section~\ref{three} for a three-channel retro-reflector.  Figure~\ref{fig12} shows the measured efficiency of back reflection (retro-reflection) when the metasurface is illuminated at $\theta_{\rm i}=0^\circ$, $ \pm28^\circ$, and $ \pm 70^\circ$. 
\begin{figure*}
	\centering
	\subfigure[]{
		\epsfig{file=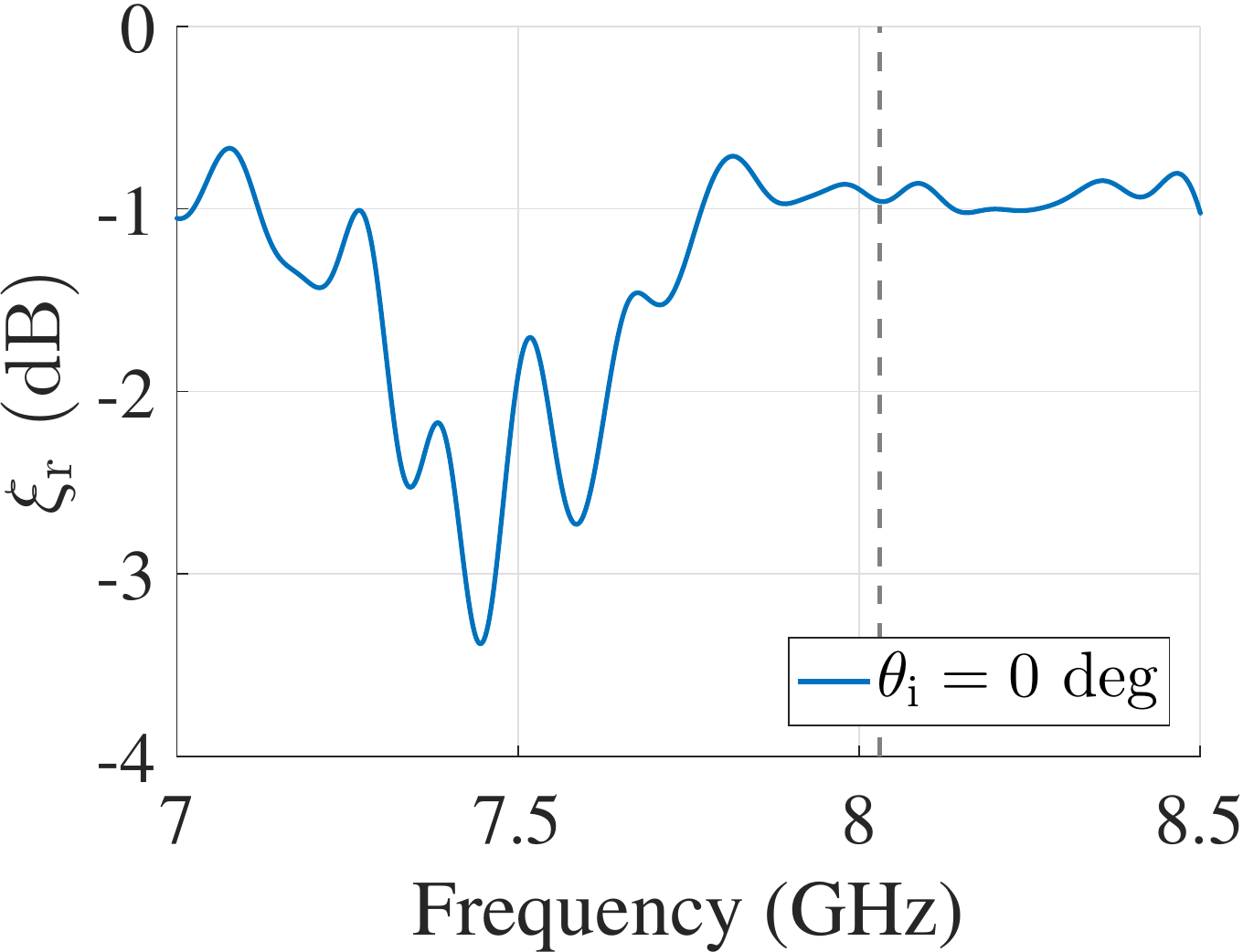, width=0.18\textwidth}  
		\label{fig12a} }
	\subfigure[]{
		\epsfig{file=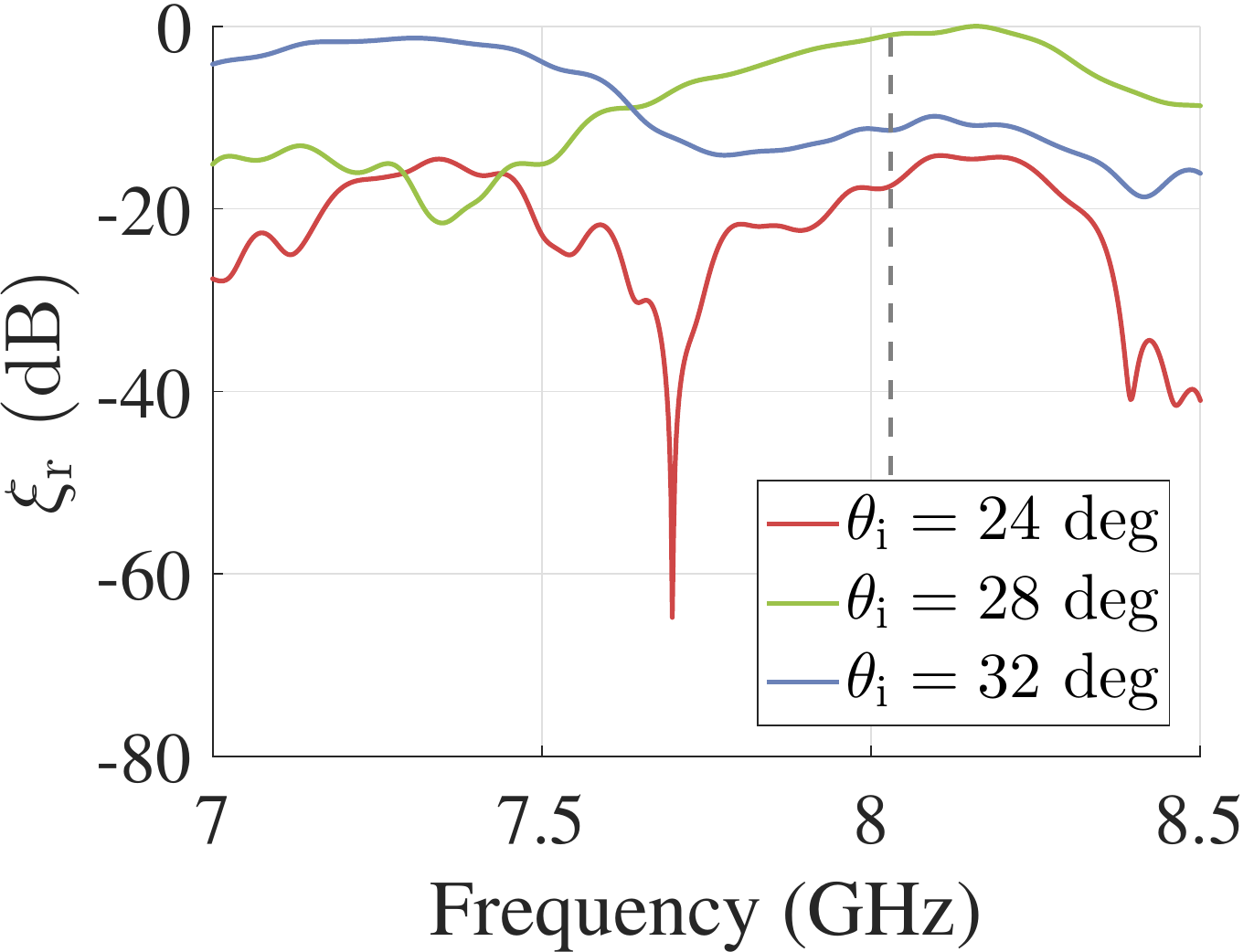, width=0.18\textwidth} 
		\label{fig12b} }  
			\subfigure[]{
		\epsfig{file=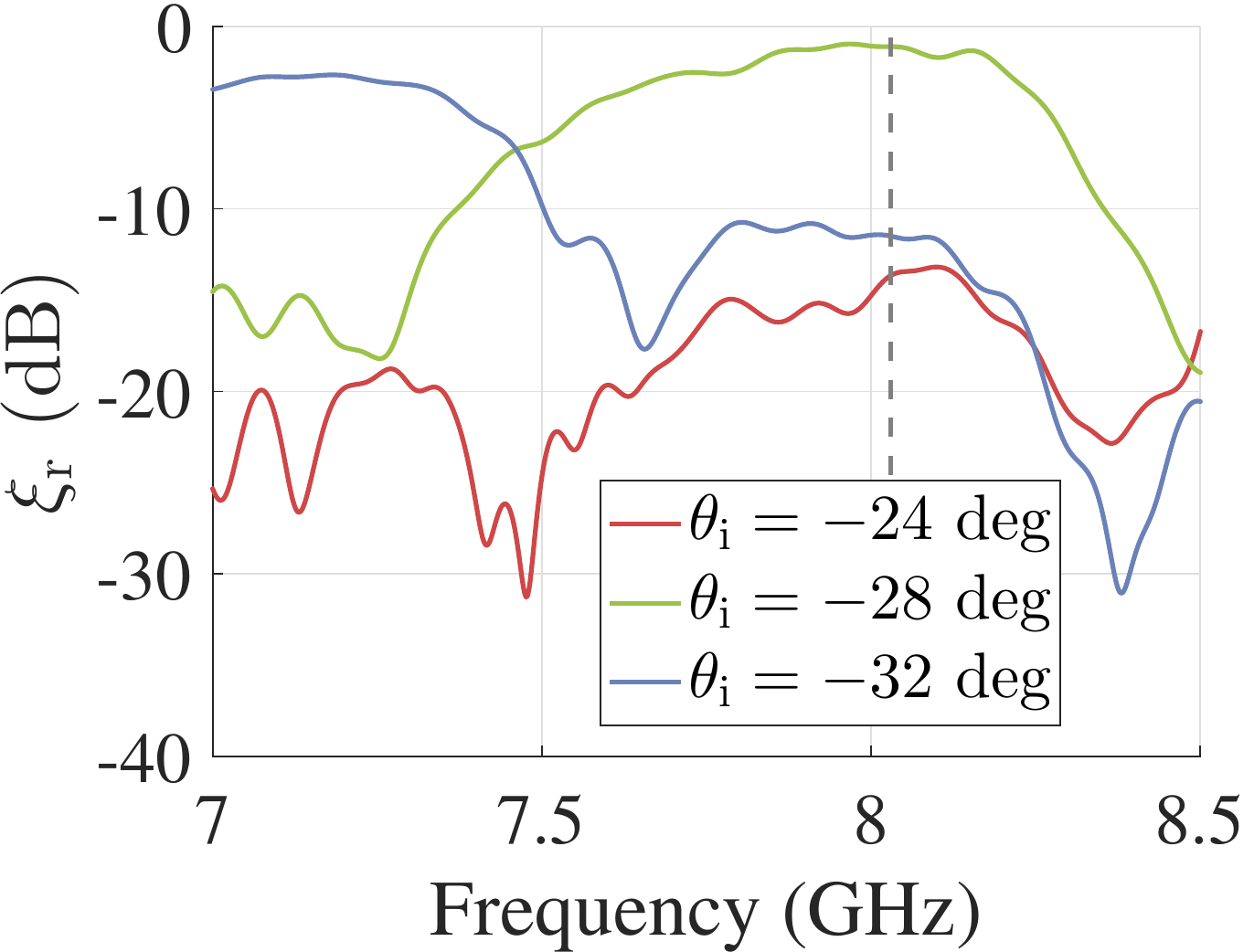, width=0.18\textwidth}  
		\label{fig12c} }
			\subfigure[]{
		\epsfig{file=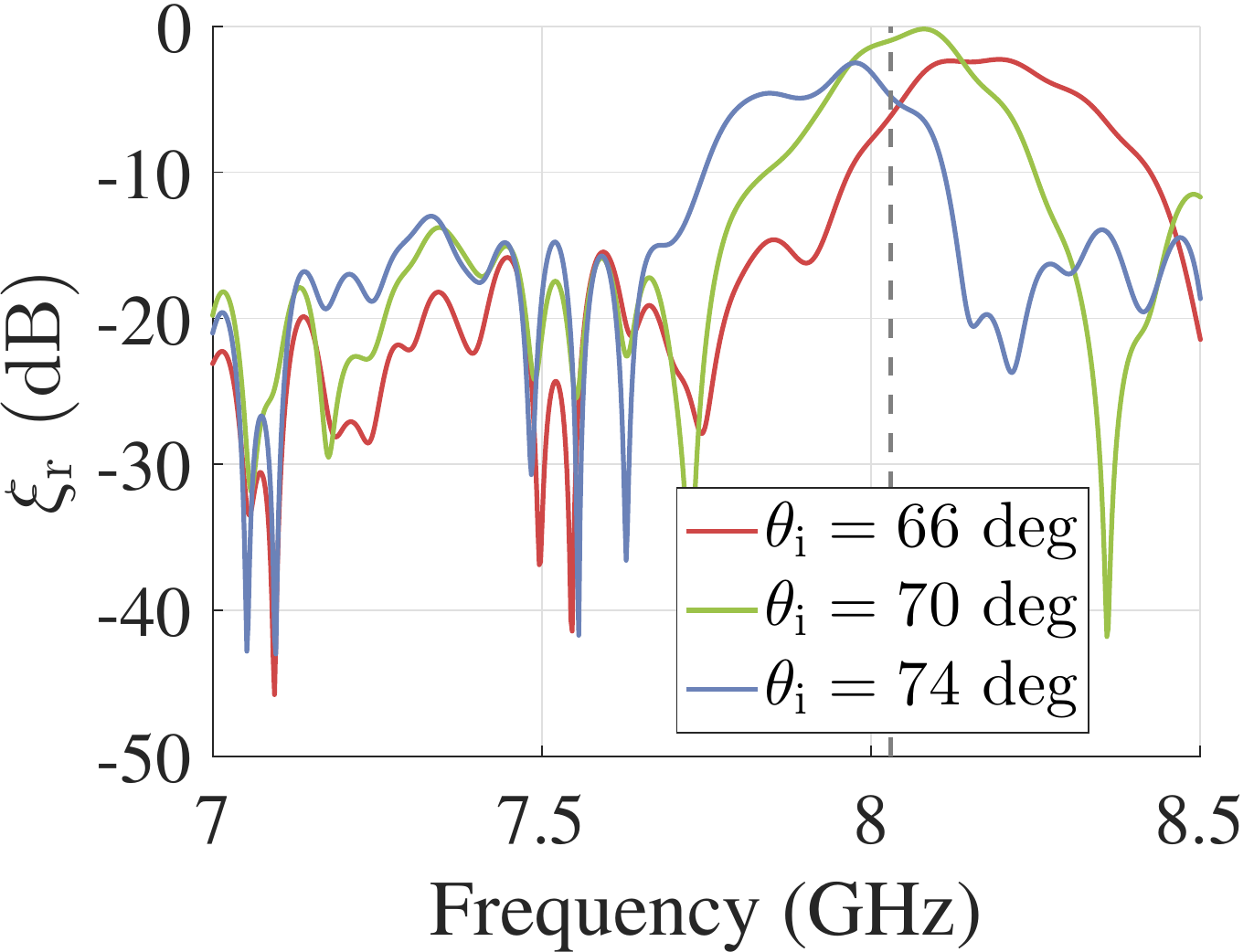, width=0.18\textwidth}  
		\label{fig12d} }
			\subfigure[]{
		\epsfig{file=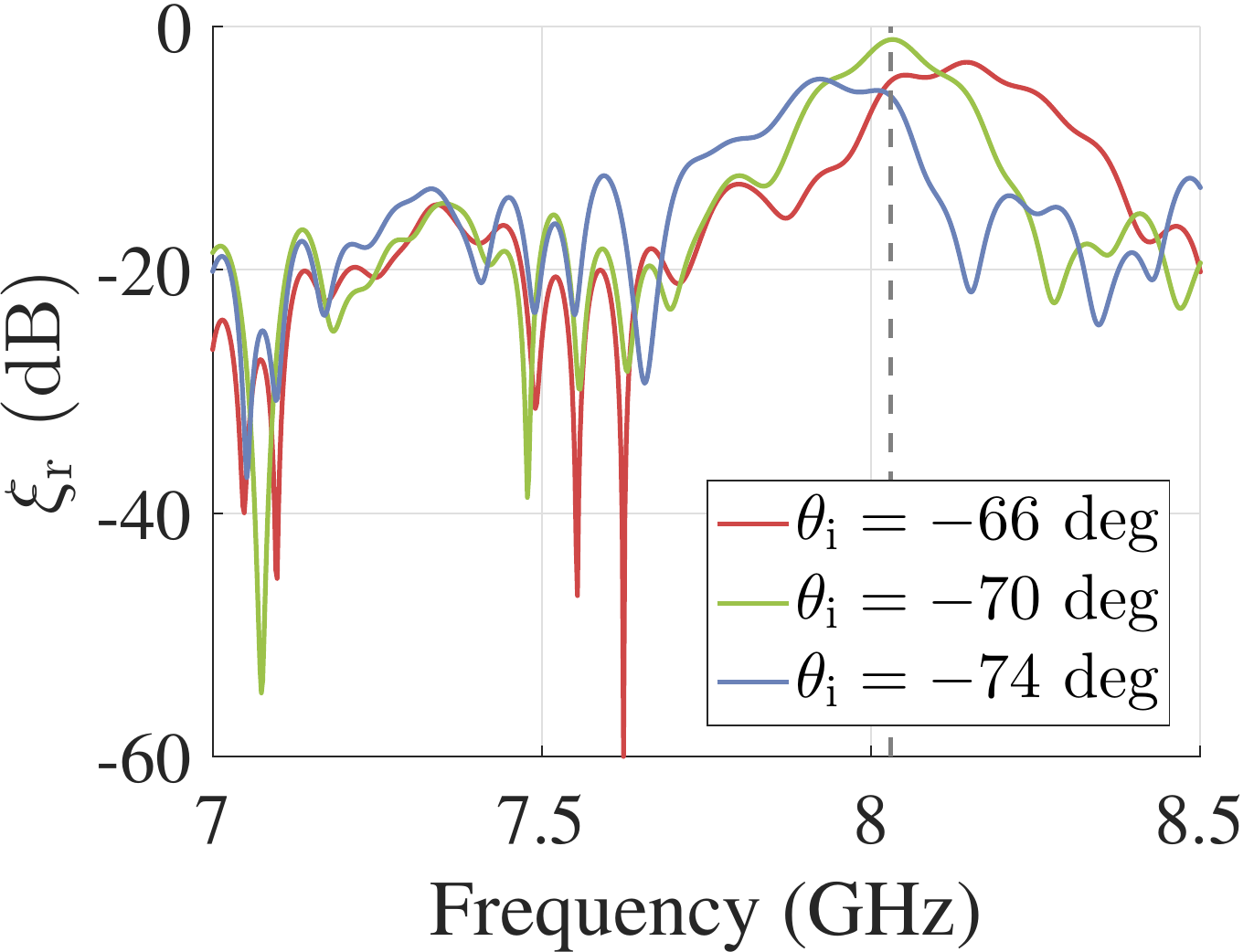, width=0.18\textwidth}  
		\label{fig12e} }
         \caption{Measured reflection efficiency of the  five-channel isolating mirror when illuminated from (a)  port~3, (b)  port~4, (c)  port~2, (d)  port~5, and (e)  port~1. The dashed line shows the operating frequency. 
	}\label{fig12}
\end{figure*}
At  the frequency of 8.03~GHz, retro-reflection in all channels is high and nearly equal. To compare the measured results at 8.03~GHz with the simulated ones, we write them in the matrix form similar to~(\ref{sm}):
\begin{equation}
(|S|^2)_{i j} = \begin{pmatrix} 
0.776  & \cdot     & \cdot  & \cdot     & \cdot     \\ 
\cdot      & 0.772 & \cdot      & \cdot & \cdot     \\
\cdot  & \cdot     &  0.802 & \cdot     & \cdot \\
\cdot      & \cdot &  \cdot     & 0.801 & \cdot     \\
\cdot      & \cdot    &  \cdot & \cdot     & 0.800 
     \end{pmatrix}.
     \label{sm2} 
\end{equation}
Here the dotes denote unknown values that were not measured in the experiment. The values at the main diagonal represent reflection efficiencies $\xi_{\rm r}$ (into the $i$-th port) expressed in the linear scale for illumination from the same $i$-th port.  As can be seen, the  simulated and measured results in (\ref{sm}) and (\ref{sm2}) are in good agreement.

\section{Discussions and conclusions}
To summarize, combining the microwave circuit and diffraction gratings theories, we proposed a concept of flat engineered multi-channel   reflectors. We revealed the existence of three basic  functionalities  available with general flat reflectors: General specular, anomalous, and   retro- reflections. 
Next, we designed three different multi-channel   devices whose response  can be described by a combination of these basic functionalities.

As  was  demonstrated, proper engineering the periodicity and surface impedance of a metasurface enables straightforward design of various multi-channel reflectors. Indeed, with the increase of the period $D$ the number of ``open'' channels grows (see Fig.~\ref{fig2b}), leading to  great design freedom. One example of possible multi-channel reflector device is a  $N$-port isolating mirror (\textit{flat} retro-reflector) depicted in Fig.~\ref{fig13a}.
\begin{figure}
	\centering
	\subfigure[]{
		\epsfig{file=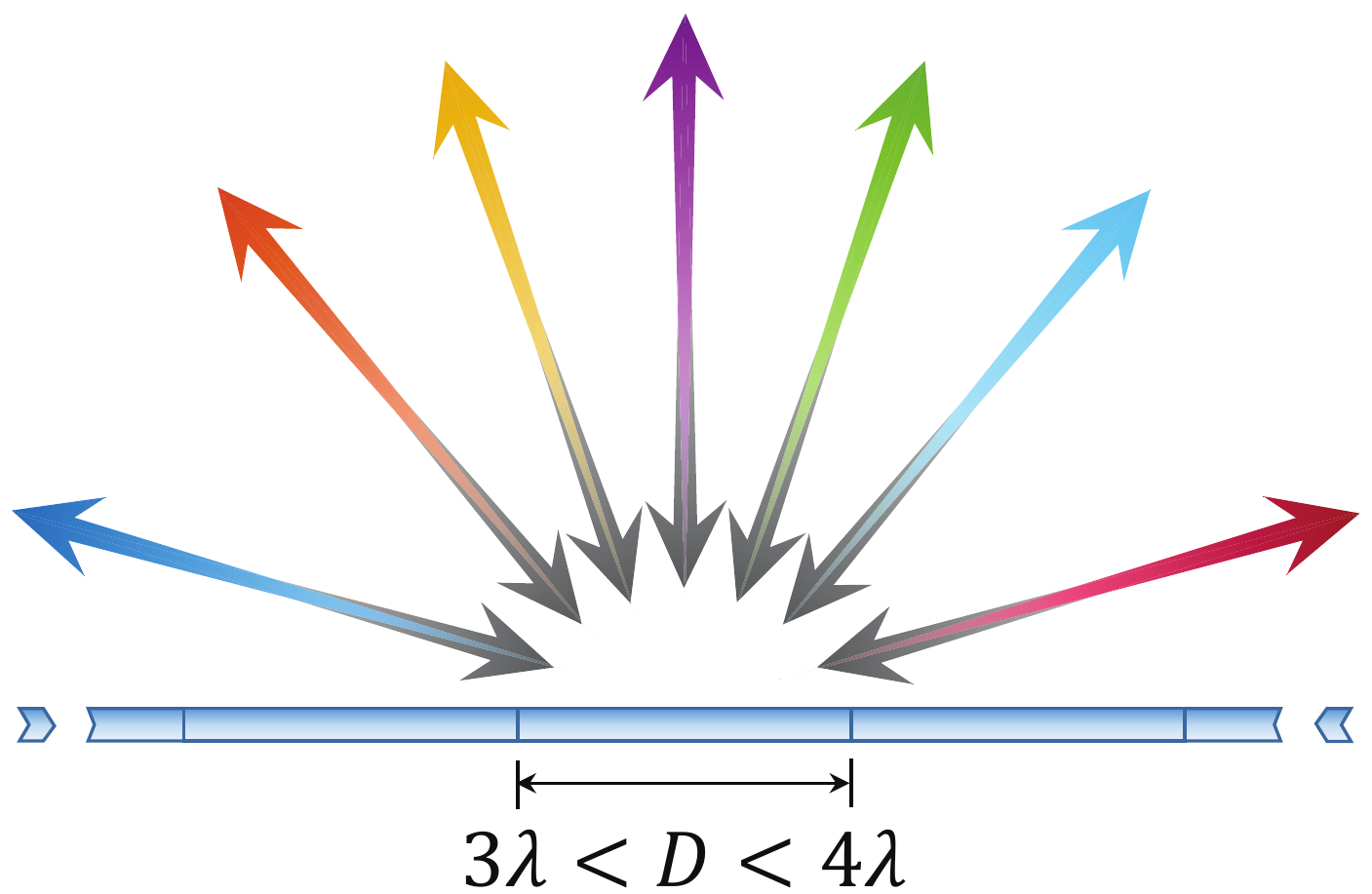, width=0.46\columnwidth}  
		\label{fig13a} }
	\subfigure[]{
		\epsfig{file=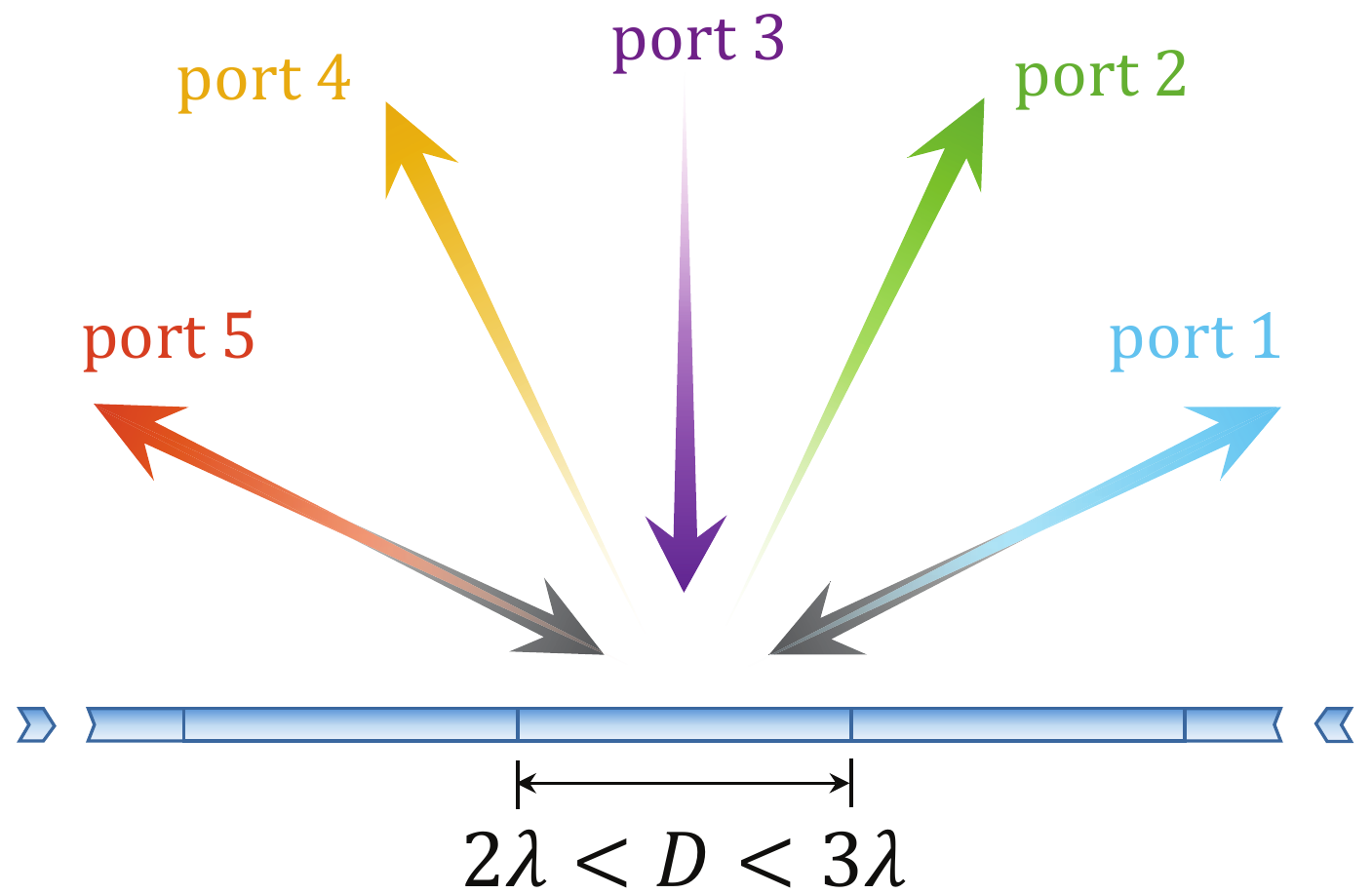, width=0.46\columnwidth} 
		\label{fig13b} }  
         \caption{(a) $N$-port isolating mirror. Wave impinging from different angles (grey arrows) is reflected back to the source (coloured arrows). (b) Multi-functional reflector. Normally incident light is split between ports~2 and~4. Ports~1 and~5 are isolated.
	}\label{fig13}
\end{figure}
All the channels (in principle, the number of channels can be very big) of this mirror are isolated from one another and, therefore, the mirror illuminated at almost any angle  would  fully reflect energy back to the source.  Such a mirror would possess unprecedented physical properties:  Observers standing around the mirror would see only images of themselves but not other observers. 
In fact, the question of what exactly an observer would see in this mirror is not straightforward and strongly depends on the number of the ``open'' channels and their isolation efficiency. 

Alternatively, large periodicity of the reflector could be used to combine different functionalities in one reflector or even to overcome the design limitations of three-channel structures. Figure~\ref{fig13b} demonstrates a beam splitter which at the same time acts as a mirror at other angles. Modifying the response from ports 1 and 5, one can greatly extend the properties which the reflector exhibits  when excited from ports 2--4. 
For the sake of simplicity, in this paper we confined our design of multi-channel reflectors to the case  when one of the channels corresponds to the normal plane-wave incidence even though the proposed metasurfaces operate also at oblique angles. The scenario with the normal incidence channel being closed further extends design possibilities, but we do not consider this case here.  

Other exciting functionalities become possible by extending  the multi-channel paradigm in three other  directions: Partially transparent films (additional channels appear in transmission), multi-channel polarization transformers, and non-periodical structures (e.g., to emulate a convex or spherical mirror response using a single flat surface).

\begin{acknowledgments}
This work was supported in part by the Nokia Foundation and the Academy of Finland (project 287894). 
The authors would like to thank Muhammad Ali and Abbas Manavi for technical help with the experimental equipment.
\end{acknowledgments}

\end{document}